\documentclass[conference,compsoc]{IEEEtran}

\usepackage{cite}
\usepackage{amsmath,amssymb,amsfonts}
\usepackage{algorithmic}
\usepackage{graphicx}
\usepackage{textcomp}
\usepackage{xspace}
\usepackage{enumitem}

\usepackage{adjustbox,multirow}

\usepackage{caption}
\usepackage[labelfont=bf,skip=5pt,figurename=Fig.]{caption}

\usepackage{subcaption}
\renewcommand{\thetable}{\Roman{table}}

\usepackage{listings}
\usepackage{xcolor}
\usepackage[scaled]{berasans}
\usepackage[T1]{fontenc}

\definecolor{light-gray}{gray}{0.95} 
\definecolor{comment-gray}{rgb}{0.41,0.6,0.33} 
\lstdefinestyle{mystyle}{
    language=Python, 
    captionpos=b,  
    basicstyle=\fontfamily{phv}\selectfont\footnotesize,  
    keywordstyle=\bfseries,  
    breaklines=true, 
    breakatwhitespace=true, 
    showstringspaces=false, 
    numbers=left, 
    numberstyle=\tiny\color{black}, 
    numbersep=10pt, 
    commentstyle=\color{comment-gray},  
    backgroundcolor=\color{light-gray},  
    xleftmargin=17pt, 
    xrightmargin=3.4pt, 
    framextopmargin=15pt,  
    framexbottommargin=10pt,  
    framexleftmargin=17pt,  
    framesep=3pt, 
    fillcolor=\color{light-gray},  
    linewidth=\linewidth, 
    basewidth=0.5em,
    deletekeywords={[2]file},
    morekeywords={match, if, then, label_host, label_file, drop, alert, allow, declassify, endorse, contains},  
    aboveskip=2ex,
    belowskip=2ex,
    escapechar=|
}

\newcommand{\incode}[1]{\lstinline{#1}}

\def\BibTeX{{\rm B\kern-.05em{\sc i\kern-.025em b}\kern-.08em
    T\kern-.1667em\lower.7ex\hbox{E}\kern-.125emX}}

\usepackage{amsmath}
\usepackage{tikz}
\newcommand*\circled[1]{\tikz[baseline=(char.base)]{
            \node[shape=circle,fill,inner sep=1pt] (char) {\textcolor{white}{#1}};}}
            
\newcommand*\circledwhite[1]{\tikz[baseline=(char.base)]{
            \node[shape=circle,draw,inner sep=1pt] (char) {#1};}}

\usepackage{filecontents}
\usepackage{booktabs}
\usepackage{float}
\usepackage{makecell}

\usepackage{graphicx}
\definecolor{strongred}{RGB}{190, 0, 0}

\usepackage{subfiles}
\usepackage{hyperref}

\usepackage[capitalize,nameinlink]{cleveref}
\hypersetup{%
	bookmarksnumbered, bookmarksopen=true, bookmarksopenlevel=1,%
}
\crefname{figure}{Fig.}{Figs.}
\crefname{listing}{Listing}{Listings}
\crefname{section}{Section}{Sections}
\crefname{table}{Table}{Tables}
\crefname{BNF}{Grammar}{Grammars}
\crefname{algorithm}{Algorithm}{Algorithms}
\crefname{line}{Line}{Lines}

\AtBeginEnvironment{appendices}{\crefalias{section}{appendix}\crefalias{subsection}{appendix}}

\usepackage{soul}

\newcommand{\cmtt}{\fontfamily{cmtt}\selectfont}

\newcommand{\tool}{\textsc{P4Control}\xspace}
\newcommand{\policy}{\textsc{NetCL}\xspace}

\renewcommand{\paragraph}[1]{\smallskip\noindent\textbf{#1.}}

\begin{document}

\title{\tool: Line-Rate Cross-Host Attack Prevention via In-Network Information Flow Control Enabled by Programmable Switches and eBPF}

\author{\IEEEauthorblockN{Osama Bajaber}
\IEEEauthorblockA{Virginia Tech \\
obajaber@vt.edu}
\and
\IEEEauthorblockN{Bo Ji}
\IEEEauthorblockA{Virginia Tech \\
boji@vt.edu}
\and
\IEEEauthorblockN{Peng Gao}
\IEEEauthorblockA{Virginia Tech \\
penggao@vt.edu}
}

\maketitle

\begin{abstract}

Modern targeted attacks such as Advanced Persistent Threats use multiple hosts as stepping stones and move laterally across them to gain deeper access to the network.
However, existing defenses lack 
end-to-end information flow visibility across hosts and cannot block cross-host attack traffic in real time.
In this paper, we propose \tool, a network defense system that precisely confines end-to-end information flows in a network and prevents cross-host attacks at line rate.
\tool introduces a novel \emph{in-network decentralized information flow control (DIFC) mechanism}
and is the first work that enforces DIFC at the network level at \emph{network line rate}.
This is achieved through: (1) an in-network primitive based on programmable switches for tracking inter-host information flows and enforcing line-rate DIFC policies;
(2) a lightweight eBPF-based primitive deployed on hosts for tracking intra-host information flows.
\tool also provides an expressive policy framework for specifying DIFC policies against different attack scenarios.
We conduct extensive evaluations to show that \tool can effectively prevent cross-host attacks in real time, while maintaining line-rate network performance and imposing minimal overhead on the network and host machines.
It is also noteworthy that \tool can facilitate the realization of a zero trust architecture through its fine-grained least-privilege network access control.

\end{abstract}

\section{Introduction}
\label{sec:intro}

Despite the dramatic growth in expenses on operational network security, we are still witnessing a rapid increase in targeted cyber attacks, such as Advanced Persistent Threats (APTs).
These sophisticated attacks often exploit multiple hosts in a network 
and laterally move to the target to access unauthorized resources or exfiltrate sensitive data~\cite{lateralmovement}.
As a result, many high-profile businesses were plagued with huge losses
\cite{breaches21st}.
These \emph{cross-host attacks} pose significant challenges to existing defenses, which lack the necessary context to correlate attack activities on different hosts and prevent attacks from damaging the network in real time.

Existing defenses treat inter-host information flows and intra-host information flows in isolation. Hence, they lack \emph{\textbf{end-to-end information flow visibility}} across multiple hosts in a network.
Network-level defenses, such as firewalls~\cite{iptable} and network intrusion detection systems (NIDSes)~\cite{snort}, have visibility into inter-host information flows between two hosts in the form of network flows (i.e., a sequence of packets sent from a source to a destination \cite{flowlabel}). However, they are unable to connect network flows to reveal cross-host attack activities due to lack of host-level visibility on intermediate hosts.
On the other hand, host-level defenses capture intra-host information flows only.
Many studies along this line employ system call monitoring to track information flows between system entities (e.g., processes, files) within a host for forensic investigation \cite{protracer2016, samuel_backtracking2003,nahid2017,fang2022back,gao2018aiql}. However, they are unable to track attack activities beyond a single host due to inadequate network-level visibility.
Although a few studies along this line proposed to associate system calls across hosts~\cite{yangji2017, yangji2018, spade2012, nahid2020},
these solutions mostly operate in post-compromise settings using historical system audit logs.
In summary, none of the existing defenses are able to block cross-host attack traffic in real time \emph{when the connection is established on the fly}.

\begin{figure*}
     \centering
     \begin{subfigure}[b]{0.49\textwidth}
         \centering
         \includegraphics[width=\textwidth]{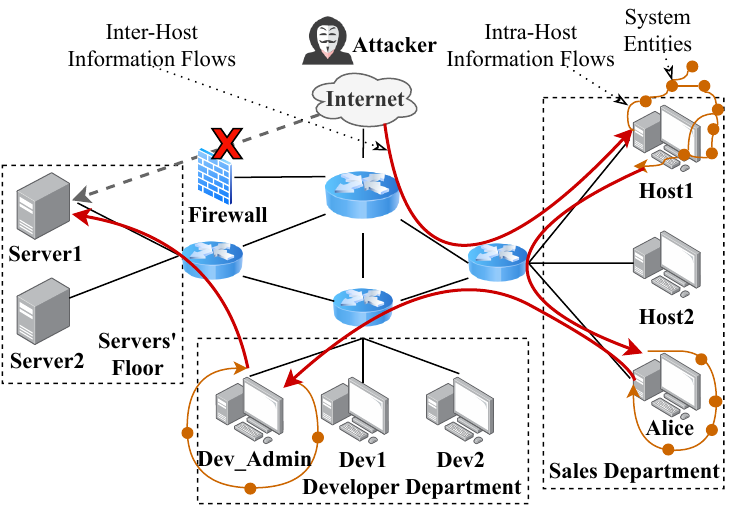}
         \caption{Motivating cross-host attack example}
         \label{fig:motivatingexample_attack}
     \end{subfigure}
     \hfill
     \begin{subfigure}[b]{0.49\textwidth}
         \centering
         \includegraphics[width=\textwidth]{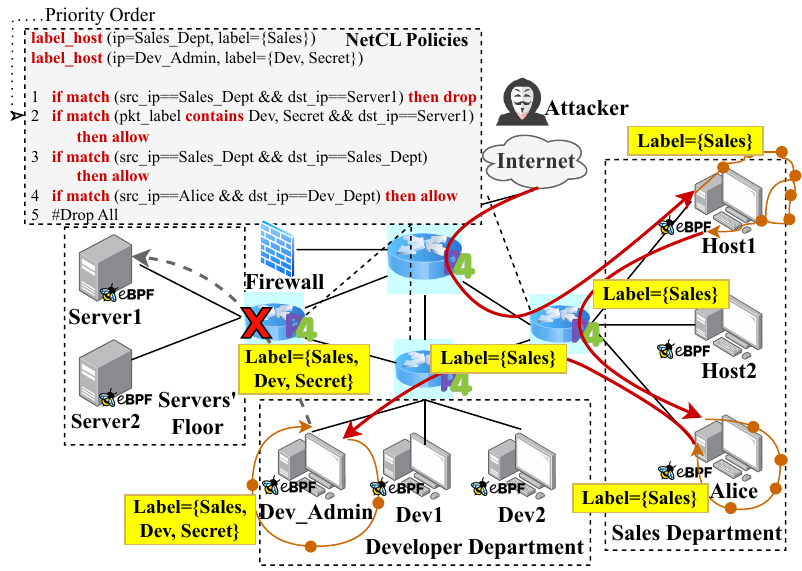}
         \caption{Defense workflow of \tool}
         \label{fig:motivatingexample_p4control}
     \end{subfigure}
        \caption{(a) A real-world attack scenario: 
        while the firewall can block any direct connections (indicated by the grey dashed arrow) from the external network to a protected server, {\cmtt Server1}, by exploiting intermediate hosts, the attacker can successfully bypass the firewall and reach its final target with four inter-host information flows (indicated by the red arrows) and multiple intra-host information flows (indicated by the orange arrows).
        (b) An illustration of the defense workflow of \tool deployed in a network of programmable switches: \tool parses DIFC labels (indicated in the yellow boxes) to precisely correlate and confine information flows across hosts and blocks cross-host attack traffic in real time based on the priority-ordered DIFC policies.}
\vspace{0ex} 
\label{fig:motivatingexample}
\end{figure*}

Consider a representative enterprise network scenario (see \cref{fig:motivatingexample_attack}). Attackers can bypass the firewall to access sensitive data on the protected host, {\cmtt Server1}, by laterally moving across four hosts with multiple inter-host information flows and intra-host information flows.

\paragraph{Goal and challenges}
To that end, the overarching goal of this work is to design and build a new network defense system that (1) enables end-to-end information flow visibility across hosts, and (2) leverages such visibility to enforce security decisions in real time to prevent cross-host attacks. 
The key challenge to achieving this goal is three-fold:

First, to enable end-to-end visibility, an effective defense must accurately correlate information flows both within and between hosts in a cross-host attack. 
Furthermore, the defense must also \emph{precisely confine} the information flow among entities (e.g., hosts, packets, processes, and files) and enforce authorized accesses.
While decentralized information flow control (DIFC)~\cite{difc1997} provides the needed fine-grained control of information flow, existing DIFC systems have only focused on operating systems (OSes)~\cite{flume2007, histar2011, asbestos2005, distributedifc2008}, distributed systems~\cite{distributedifc2008, distributedifc2012}, and cloud computing~\cite{cloudifc2012,cloudifc2013}. 
None of them have enforced DIFC \emph{at the network level}.

Second, enforcing DIFC at the network level is highly challenging due to the huge volume of traffic in enterprise networks. An effective defense must be able to correlate information flows and enforce DIFC policies on the fly, without imposing significant overhead on the network performance. The defense must be seamlessly integrated with the existing network infrastructure and must not affect the \emph{line-rate processing} of large amounts of 
benign traffic.

Third, human analysts with domain knowledge are crucial for defenses~\cite{gao2018aiql}.
An effective defense must hide the complexity of low-level DIFC enforcement and allow the network administrator to tailor the defenses for different attacks, through a flexible and expressive 
policy interface.

\paragraph{Contributions}
We propose \tool, a network defense system that precisely confines end-to-end information flows and prevents cross-host attacks in real time when the connection is established on the fly. \tool introduces a novel \emph{\textbf{in-network DIFC mechanism}} for precise information confinement and line-rate DIFC policy enforcement. The mechanism includes a secure network-level DIFC model with a category system and an in-network DIFC enforcement approach enabled by programmable switches and eBPF. \tool also provides a flexible and expressive policy framework to specify a wide range of DIFC policies.

\tool creates DIFC labels for network entities (e.g., hosts and packets) and system entities (e.g., processes and files) and propagates these labels along intra-host and inter-host flows between the entities. These labels can encode different \emph{categories, secrecy and integrity levels, etc.} \tool then enforces DIFC policies specified by the network administrator on labeled flows in the \emph{network data plane} at line rate. In addition, \tool provides (1) safe mechanisms to declassify secret data to authorized readers or endorse data as high integrity, and (2) a tainting mechanism for fine-grained tracking of the propagation path of a sensitive file in the network to limit its reachability. Our network-level DIFC model formalizes these operations.

\tool uniquely leverages the emerging programmable switches and eBPF to realize the DIFC model in the data plane. 
Programmable switches offer data-plane programmability through P4 \cite{bosshart2014} and guarantee customized Tbps line-rate packet processing. 
This enables \tool to process labeled network traffic and enforce line-rate DIFC policies. To address the key challenge of limited switch memory and minimize the network overhead, \tool employs a secure-yet-practical in-network DIFC enforcement approach with \emph{tailored techniques} to label network traffic, match DIFC policies, enforce per-flow decisions, and enable declassification/endorsement controls.

To propagate DIFC labels within each host and from/to the network, we develop a lightweight host agent based on the eBPF 
technology \cite{ebpf}. 
Our host agent is lightweight and readily deployable \emph{without any kernel modifications}. This differentiates our approach from previous DIFC works~\cite{asbestos2005, oconnor2018, histar2011}, which require extensive OS kernel modifications to track intra-host information flows.
Our host agent enables \emph{lightweight DIFC label persistence} by attaching carefully defined eBPF hooks in the kernel to capture the complete chain of intra-host system events and accurately propagate DIFC labels, with minimal overhead on the host machine.

While programmable switches enable in-network DIFC, it is challenging for network administrators to directly program the data plane using P4, which is low-level and can be error-prone~\cite{lucid2021}. To unlock the powerful in-network DIFC context, we design \emph{Network Control Language (\policy)}, an expressive domain-specific language that enables the network administrator to specify DIFC policies that match cross-host flows and trigger a wide range of defense actions, including preventing data exfiltration, detecting unauthorized access, declassifying information, and limiting the reachability of sensitive files and the spread of malware.
\policy policies follow a priority-based enforcement similar to the traditional firewall policies. To further enhance the defense agility against attacker's possibly changing strategies, \tool employs an efficient compilation mechanism that supports \emph{dynamical update} of \policy policies at runtime without interrupting network traffic.

\paragraph{Evaluation}
We extensively evaluated \tool's defense effectiveness and coverage, scalability, system capacity, and system overhead, and compared with firewall (iptables~\cite{iptable}), NIDS (Snort~\cite{snort}), and SDN-based IFC solutions (PivotWall~\cite{oconnor2018}). 
We deploy \tool on both a physical testbed (with a Tofino programmable switch~\cite{tofino}) and representative large enterprise topologies constructed using a packet-level simulation, 
and use synthetic and real-world enterprise traces.
The evaluation results demonstrate that:
(1) \tool successfully prevents a wide range of stealthy cross-host attacks 
that all evade firewall and NIDS, while adding only a negligible $\sim$110~ns overhead to the packet processing time.
(2) \tool successfully prevents real-world enterprise cross-host attacks from the DARPA OpTC dataset~\cite{darpa} and the LANL Unified Host and Network dataset~\cite{lanl}, while achieving 99.9~Gbps throughput on the 100~Gbps (per-port) programmable switch, incurring minimal overhead on benign traffic.
(3) \tool is robust against control plane attacks and achieves robust performance even with an attack strength of 1 million packets/second, compared to
PivotWall, which 
fails to install 99\% of legitimate connections.
(4) The eBPF-based agent adds an overhead of 1-7 ms to the total time of the monitored system calls, imposing minimal overhead on the host performance.

Clearly, these results demonstrate that \tool outperforms existing defenses in combating cross-host attacks by a significant margin.
\tool's in-network DIFC approach based on programmable switches and eBPF offers robust defensive effectiveness and wide attack coverage, maintains line-rate performance, and imposes minimal overhead on the network and host machines.
We open-source the prototype of \tool at~\cite{sourcecode}.

\paragraph{Significance of the work}
\tool is the first work that enforces DIFC at the network level at line rate. \tool is also the first work that uses programmable data planes to enforce complex secrecy and integrity policies at line rate. \tool introduces a new paradigm of network-level APT defenses using programmable data planes, which differs from all existing system-level APT defenses based on system audit logs and system provenance graphs (e.g., \cite{protracer2016, samuel_backtracking2003,nahid2017,fang2022back,gao2018aiql,yangji2017, yangji2018, spade2012, nahid2020}). 
\tool can be seamlessly integrated into the existing network infrastructure with minimal modifications and overhead, transforming it into a defense backbone.
\tool radically shifts from the traditional ``castle-and-moat'' security model
that relies on perimeter defenses and implicit trust inside the network, and aligns with the principles of zero trust. 
By precisely confining information in a network with DIFC labels, \tool can facilitate the realization of a 
\emph{zero trust architecture}~\cite{zerotrust} through its fine-grained least-privilege network access control.

\section{Background and Motivating Example}
\label{sec:motivation_background}

\paragraph{Motivating example}
Organizations face challenges in defending their resources against cross-host attacks like APTs, particularly as networks become larger and more complex~\cite{lateralmovement}.
Consider an enterprise network (see~\cref{fig:motivatingexample_attack}).
Hosts and servers across different departments are interconnected, each protected by a perimeter firewall. 
Alice, a former developer, 
transitions to a new role in the Sales Department, while retaining access to the Developer Department to complete a project.
Users like Alice (i.e., those with multi-domain access) are common in enterprise networks.

An external attacker aims to compromise the integrity of {\cmtt Server1} and exfiltrate data.
To protect {\cmtt Server1}, the firewall is configured to only allow direct connections between {\cmtt Dev\_Admin} and {\cmtt Server1}. If the attacker attempts to directly connect to {\cmtt Server1} from the external network, the attempt will be blocked (i.e., grey dashed arrow).
However, the attacker can target {\cmtt Alice}'s dual access as a \emph{stepping stone}: 
(1) infiltrating {\cmtt Host1} by exploiting a zero-day vulnerability through spear-phishing; (2) pivoting connections and gaining access to {\cmtt Alice}, {\cmtt Dev\_Admin}, and {\cmtt Server1} in sequence; (3) launching a ransomware attack (compromising integrity) or data exfiltration (compromising confidentiality) on {\cmtt Server1}.
The attack path consists of four inter-host flows (i.e., red arrows) and multiple intra-host flows (i.e., orange arrows).
Since firewalls can only observe coarse-grained information (e.g., IP addresses and port numbers) and use such limited information to block \emph{direct} communications between two hosts, they miss such stepping-stone attacks.

\begin{figure}[t]
    \centering
    \includegraphics[width=\linewidth]{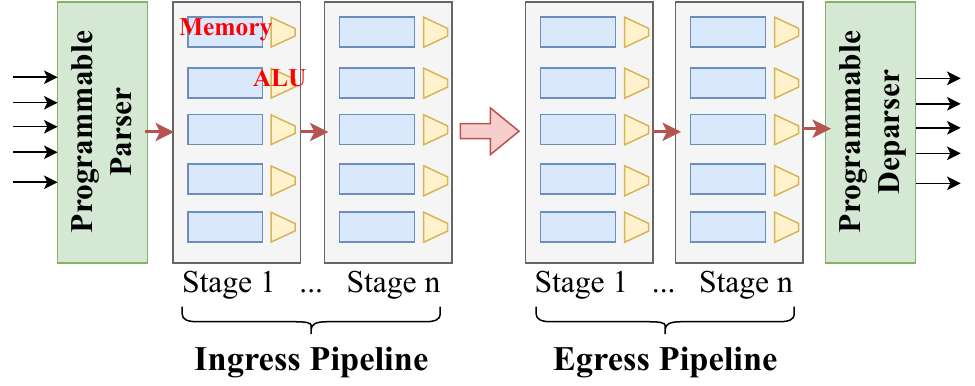}
    \caption{Protocol independent switch architecture}
    \label{fig:tofinoarch}
    \vspace{-2ex}
\end{figure}

\paragraph{Programmable data plane as a defense solution} 
Programmable switches have recently attracted increased attention for their data-plane programmability that achieves line-rate performance with low overhead, compared to the software-defined networking (SDN) counterparts that often require extensive control-plane communications.
These switches can be programmed using P4~\cite{bosshart2014} for customizing packet processing (thus also called P4 switches). 
As long as a P4 program can be successfully compiled, the data plane guarantees to process packets at \emph{Tbps line rate}~\cite{tofino}.

\begin{figure*}[t]
\centering
\includegraphics[width=\linewidth]{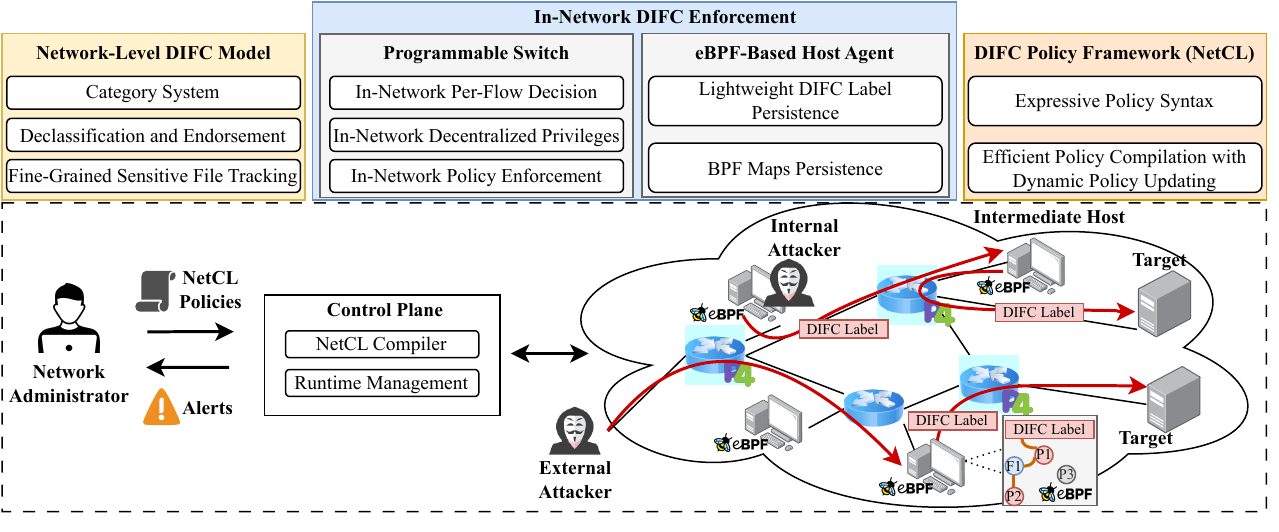}
\caption{Overall architecture of \tool} 
\label{fig:systemarch}
\vspace{-1ex}
\end{figure*}

\cref{fig:tofinoarch} illustrates the switch architecture. 
P4 programs specify the packet headers 
and the operations on these headers. 
The programmable parser parses the user-defined packet headers. 
These headers go through multiple hardware stages with Arithmetic Logic Units (ALUs) and match/action tables, where match fields and types (e.g., exact/range/ternary matching) can be specified.
These stages use Static RAM (SRAM) and Ternary CAM (TCAM) for match lookups. 
While SRAM supports exact matching and persists data across packets for stateful processing, TCAM supports wildcard matches over header fields.
Unfortunately, \emph{switching ASICs only offer limited memory} (e.g., hundreds of MB of SRAM and tens of MB of TCAM)~\cite{tofino}. 
To guarantee line-rate processing, P4 programs limit the operations in each stage. 
When the packet header matches a table, it triggers an action. 
The programmable deparser reassembles the packet before it leaves the switch and gets forwarded.

Recent works proposed different implementations inside the programmable switch to defend against distributed denial-of-service (DDoS)~\cite{poseidon2020,jaqen2021}, link flooding~\cite{Jiarong2021,mew2023}, covert channels~\cite{Jiarong2020}, and BYOD security~\cite{qiao2020}. 
However, none of these works have focused on cross-host attacks.

\paragraph{DIFC}
Information flow control (IFC) monitors and regulates the movement of information within a system.
In classical centralized IFC~\cite{blp1976, biba1976,denning1976lattice}, a central authority assigns predefined security labels to subjects (e.g., processes) and objects (e.g., files) and enforces IFC policies. Only the central authority can change labels or policies.
Decentralized IFC (DIFC)~\cite{difc1997} is a generalization of the classical IFC and offers more flexibility and autonomy. 
In DIFC, data owners can set their own security policies and labels for their data, and grant permissions to subjects to alter their labels. 

While prior works have integrated DIFC with OS~\cite{flume2007,asbestos2005,histar2011} and extended it to Android~\cite{weir2016} and distributed systems~\cite{distributedifc2008, distributedifc2012}, these implementations require substantial modifications to the kernel or user-space application, which is highly complex. Besides, their enforcement happens within individual hosts rather than in the network. These DIFC systems also incur a significant
overhead on system operations (e.g., 30-40\% slower in the Flume system~\cite{flume2007}) due to the low processing power of host systems, failing to meet our line-rate requirement.
While other works have employed IFC in cloud environments~\cite{cloudifc2012,cloudifc2013}, 
these works focus on the interactions between users and cloud providers, which are distinct from network communications.

\section{\tool System Overview}
\label{sec:overview}

We present the overall architecture of \tool in \cref{fig:systemarch} and briefly describe its defense workflow below. 

\emph{Initialization:}
(1) The network administrator specifies \policy policies for assigning DIFC labels (i.e., a set of DIFC tags) to hosts and matching network flows. 
(2) The programmable switch
sends a control packet containing the specified DIFC label to the corresponding eBPF-based host agent.
(3) The host agent initializes existing processes and files in the host with the received DIFC label.

\emph{DIFC context persistence:}
(4) As an attacker enters the host system, the host agent propagates the DIFC label between system entities along with the attacker's activities. When the attacker pivots to another host, the host agent propagates the DIFC label to the network by incorporating the label into the
outgoing network flow (in a customized DIFC packet header).
(5) The host agent on the receiver host extracts the DIFC label from the network flow, merges it with the label of the receiver process, and continues propagating the updated label. This way, the host agents maintain the DIFC context persistence across the network.

\emph{DIFC policy enforcement:}
(6) \policy policies specified by the network administrator are compiled into different in-network policies, which are inserted into the programmable switch's match/action tables.
(7) When a labeled network flow arrives at the switch, the switch extracts DIFC tags from the DIFC label and uses these tags to correlate all previous flows. The switch then matches these tags against the in-network policies to trigger the corresponding security action (e.g., drop the flow) at line rate.

\cref{fig:motivatingexample_p4control} illustrates how \tool counters the attack scenario described in \cref{sec:motivation_background}. After initialization, the processes on {\cmtt Dev\_Admin} have the label \{\texttt{Dev,Secret}\} with the tags \texttt{Dev} and \texttt{Secret}. When the attacker pivots from the Sales Department to {\cmtt Dev\_Admin}, the \{\texttt{Sales}\} label is propagated to the network flow, and the attacker process on {\cmtt Dev\_Admin} then has the label \{\texttt{Sales,Dev,Secret}\}, indicating that the process has previous interactions with entities that have the \texttt{Sales} tag. Subsequently, when the attacker tries to connect to {\cmtt Server1}, the programmable switch detects the presence of the \texttt{Sales} tag in the network flow and realizes that the flow traverses the Sales Department. The switch then enforces the matched policy that has the highest priority (i.e., the {\cmtt drop} policy) to drop the traffic.

\paragraph{Threat model}
Our threat model is similar to that of many previous works on programmable switch-based network defenses~\cite{poseidon2020,jaqen2021,Jiarong2021,mew2023,Jiarong2020,qiao2020}
and host-level auditing~\cite{protracer2016, samuel_backtracking2003,nahid2017,fang2022back,gao2018aiql,yangji2017,yangji2018,spade2012,nahid2020}.
We assume the presence of an attacker seeking to access or modify unauthorized resources, exfiltrate confidential data, or spread malware, either from within the network or externally, by exploiting trust relationships among networked hosts.
Our trusted computing base includes programmable switches, the control plane, and host agents. We assume that the OS kernels are secure from compromise, and that the network administrator specifies policies correctly especially regarding declassification. We do not consider malicious administrators who can disable the host agent or tamper with DIFC labels, or implicit flows like covert and timing channels.
Tamper-proof and tamper-evident auditing techniques~\cite{paccagnella2020custos,ahmad2022hardlog}
can be leveraged to further secure our host agents.

\section{Network-Level DIFC Model}
\label{sec:in-network_model}

\tool implements a secure network-level DIFC model with a category system to associate entities with DIFC labels, a declassification and endorsement mechanism, and a mechanism for fine-grained tracking of sensitive files.

\subsection{Category System}
\label{subsec:tags}

Our category system associates different network entities (e.g., hosts and packets) and system entities (e.g., processes and files)
with DIFC tags and DIFC labels.
We extend the Flume DIFC model~\cite{flume2007}, a host-level DIFC model, to the network level and \emph{inherit} its security guarantees.
As in Flume, \tool uses DIFC tags to govern the flow of information between processes and files residing on the same machine and processes residing on different machines. Tags are assigned to both subjects (processes) and objects (files). Directories are treated as files. A set of tags form a DIFC label
\cite{denning1976lattice}.
These tags and labels can encode various categories (e.g., different enterprise departments) and secrecy and integrity levels (e.g., top-secret, secret, and unclassified) for entities to achieve enhanced multi-level security, adhering to the principle of least privilege~\cite{saltzer1975}.

Let $S_p$ and $I_p$ be the secrecy and integrity labels of entity $p$, respectively, and let $L_p=S_p \cup I_p$ be its overall label. Following Flume's \emph{safe message rules}, process $p$ can send a message to process $q$ only if $S_p \subseteq S_q$ (i.e., ``no read up, no write down''~\cite{blp1976}) and $I_p \supseteq I_q$ (i.e., ``no read down, no write up''~\cite{biba1976}).
To extend the label visibility from a single host to the network, \tool incorporates the \emph{labeling of network packets}. When a message $m$ is sent from process $p$ to process $q$, a label $L_m$ is assigned to $m$. For the message $m$ to be delivered to $q$, it must satisfy the condition $L_p \subseteq L_m \subseteq L_q$ before delivery.

To comply with the safe message rules, processes must change their labels before they can communicate with other processes or files.
For example, a process that carries a \texttt{Sales} tag can only share data with processes having a matching \texttt{Sales} tag.
Note that in Flume~\cite{flume2007}, explicit label change requires the prediction of communication patterns of subject processes to adjust labels. However, this approach is impractical in unpredictable environments and requires significant effort to modify all applications' code, limiting the DIFC's efficacy.
Therefore, \tool adopts implicit label change as in Asbestos~\cite{asbestos2005}, allowing \emph{implicit label propagation} between processes and files:
If process $p$ communicates with process $q$, then both of their labels merge to update $L_q$ (i.e., $L_q = L_p \cup L_q$). If processes $p$ and $q$ are on different machines, $p$ appends its label to the outgoing packets, which is then propagated to process $q$ upon arrival. 
For files, if a process $p$ reads from an existing file $f$, it initiates a flow from $f$ to $p$, propagating $f$'s label to $p$. This confirms that $p$ has accessed data tagged with $L_f$. When $p$ writes to a new file $f$, $p$ specifies $L_f$ for $f$, which includes all tags in $L_p$. This design is vital in tracking long-going attacks that involve data theft stored for future exfiltration.

\subsection{Declassification and Endorsement}
\label{subsec:privilege}

Our model supports decentralized privileges to declassify (remove secrecy tags) or endorse (add integrity tags) information. Each tag $t$ has two associated capabilities: $t^+$ allows a process to add tag $t$ to its label, and $t^-$ allows removing tag $t$. 
Let $C_p$ be the set of capabilities that process $p$ has.
Process $p$ can add (or remove) tag $t$ to its label only if it has the capability $t^+ \in C_p$ (or $t^- \in C_p$).
For secrecy, the capability $t^-$ allows a process to declassify information associated with tag $t$. For integrity, the capability $t^+$ allows a process to endorse its state with an integrity level associated with the tag $t$. As remote hosts are untrusted, they are modeled as an untrusted process $x$ with an empty label (i.e., $L_x = \{\}$). Therefore, to interact with the outside world, a process must have the capability to reduce its label to $\{\}$.

\subsection{Fine-Grained Tracking of Sensitive Files}
\label{subsec:file_tracking}

The model we have described does not provide enough granularity to track individual high-value files.
When a sensitive file is declassified, it is hard to regulate its accessibility to unauthorized readers.
To address this, we further enhance our DIFC model with a special \texttt{TrackerID} tag and a \emph{tainting mechanism} for specific files.
If a process reads a tagged file, it inherits the \texttt{TrackerID}, which is then propagated to other processes and to the network when the file data is exported.
This alerts the programmable switch that a file with the \texttt{TrackerID} tag is being transmitted. 
This design offers two significant benefits. 
First, \texttt{TrackerID} enables fine-grained tracking and policy enforcement on specific sensitive files. 
Second, we can monitor \texttt{TrackerID} to create a provenance graph, which is useful for tracking declassified sensitive files and forensic analysis.

\section{In-Network DIFC Enforcement} \label{sec:in_network_realization}

\tool realizes the network-level DIFC model in the data plane:
\tool leverages eBPF to realize the implicit label propagation within each host and between hosts to maintain the DIFC context persistence.
\tool leverages programmable switches to further \emph{regulate} the label propagation between hosts 
by parsing the DIFC label carried in the network flow and enforcing line-rate DIFC policies.  
To minimize the network overhead, \tool employs an in-network per-flow decision mechanism that enforces DIFC policies at the flow granularity, removing the need for labeling and matching every packet in the flow. 
\tool also employs a multi-table flow matching technique to support a large number of in-network policies with limited switch memory.

\subsection{In-Network Per-Flow Decision}
\label{subsec:per_flow_decision}

\begin{figure}[t]
\centering
\includegraphics[width=0.93\linewidth]{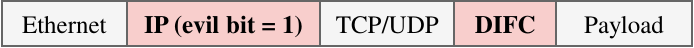}
\caption{Headers of DIFC-labeled network packet} 
\vspace{-1ex}
\label{fig:packet}
\end{figure}

To carry the DIFC label in network traffic, \tool employs a customized network packet format (see \cref{fig:packet}). 
We set the reserved bit in the IP fragment field (known as the ``evil'' bit) 
to distinguish labeled packets from regular packets and use a \emph{DIFC packet header} to carry DIFC tags. 
These tags can be extracted by the programmable switch.
Our implementation considers a 32-byte DIFC packet header, which supports 256 distinct tags (each bit represents a tag).
This indicates the number of categories and security levels in the network that can be supported by the Tofino 1 switch model (Tofino 1 model supports a maximum of 256-bit matching keys within TCAM).
It is noteworthy that this capacity largely exceeds the U.S. Department of Defense minimum access control requirement of 16 sensitivity classifications and 64 categories~\cite{tcsec}. The latest switch models (e.g., Tofino~2/3~\cite{tofino2,tofino3}) have 3$\times$ more resources than Tofino 1 and can support a larger number of tags.

\paragraph{Per-flow decision} 
A naive way of carrying DIFC labels is to label every packet in a network flow. However, this would waste resources, as the same security decision applies to all packets in the same network flow.
To reduce the network overhead, \tool employs an \emph{in-network per-flow decision} mechanism using stateful registers in programmable switches.
Rather than labeling every packet in a network flow, \tool only adds the DIFC packet header to the \emph{initial packets} of a flow. 
The security decision for the flow is then maintained in a match/action table (called {\cmtt ConnDec} table), which includes the flow's 5-tuple key ($\mathrm{IP_{src}}$, $\mathrm{Port_{src}}$, $\mathrm{IP_{dst}}$, $\mathrm{Port_{dst}}$, $\mathrm{Protocol}$) and the decision value. 
Subsequent packets in the flow will match the corresponding entry in the {\cmtt ConnDec} table, and the same decision will be applied.

\tool employs different strategies to \emph{support different network protocols}.
For TCP connections, the host agent adds the DIFC packet header to the SYN packet during the three-way handshake. This guarantees that the label is received by the switch for successful connections.
However, in UDP connections, where packet delivery is not guaranteed, the switch may not receive the packet that carries the label. To address this, the host agent adds the DIFC packet header to the first few UDP packets in a new connection. Once the switch receives a packet with the DIFC packet header, it crafts an ACK packet using the hardware packet generator and sends it back to the sender host. This acknowledges that the DIFC label has been received, allowing the host to send the remaining packets without additional DIFC packet headers.
For ICMP, the host agent adds the DIFC packet header to the request and reply ICMP packets.

\paragraph{Hardware decision buffering}
An issue arises when relying solely on the {\cmtt ConnDec} table. 
Although match/action tables can handle a large number of entries, the control plane must be involved to add entries to {\cmtt ConnDec} for every new network flow.
This means that the switch has to request the control plane to install an entry after matching a new flow, introducing a delay, known as round-trip time (RTT), from when the switch matches a decision for a new flow to the point where the entry is inserted into {\cmtt ConnDec}.
During this time period, the remaining packets of the matched flow can arrive at the switch before their entry is inserted.

To address this issue, we implement a \emph{hardware buffer} structure using stateful registers, which can be directly updated by the switch's data plane at line rate.
When a new network flow arrives, \tool matches the flow using the DIFC label in the flow's first packet, inserts an entry to the buffer on the fly, and sends a request to the control plane to update {\cmtt ConnDec}. Each entry in the buffer stores the CRC hash value of the flow's 5-tuple key and the security decision. 
When the remaining data packets arrive at the switch, \tool calculates their hash values and matches them with the buffered decision until the corresponding entry is inserted into {\cmtt ConnDec}.

Note that hash collisions can happen. 
If a new network flow, {\cmtt flow2}, has a  collision with an existing flow, {\cmtt flow1}, in the buffer, \tool evicts {\cmtt flow1}'s entry to make room for {\cmtt flow2}.
However, {\cmtt flow1}'s entry might be evicted before its corresponding entry is inserted into {\cmtt ConnDec}.
This can happen if {\cmtt flow2} and {\cmtt flow1} arrive at the switch within a very short time (i.e., RTT) and have the same hash key {\cmtt h}. 
To address this issue, \tool recirculates the remaining packets of {\cmtt flow1} for a time exceeding the expected RTT to ensure that {\cmtt flow1}'s entry is inserted into {\cmtt ConnDec}. 

However, it is worth noting that such recirculation rarely happens as {\cmtt flow2} and {\cmtt flow1} need to (1) have a collision, and (2) arrive at the switch within RTT (typically in milliseconds). 
Otherwise, {\cmtt flow1}'s entry is already inserted into {\cmtt ConnDec}, and thus, the remaining packets of {\cmtt flow1} can bypass the buffer checking, and its entry in the buffer can be safely evicted.
Our current implementation uses a buffer with up to $2^{32}$ entries, utilizing the output of CRC-32 hash function as the key. 
The latest Tofino~2 hardware~\cite{tofino2} has more resources and can accommodate a buffer with up to $2^{64}$ entries, further reducing the chance of collisions.

\paragraph{Mitigating flooding attacks}
Although \tool's in-network defense effectively shields against cross-host attacks, it remains critical to mitigate the potential risk of exploitation posed by malicious hosts.
For example, the attacker can exhaust the stateful storage of the {\cmtt ConnDec} table by initiating many new connections. 
To counter this, \tool employs a \emph{rate-limiting} strategy that restricts the number of requests from an IP address over a certain period. 
\tool also periodically removes inactive connections from {\cmtt ConnDec} to avoid resource exhaustion.

\subsection{In-Network Decentralized Privileges} 
\label{subsec:in-network_decentrlized_privileges}

We now describe how \tool realizes the decentralized privileges capabilities.
Note that \tool does not focus on regulating communications within hosts, which has been extensively studied in existing OS-level DIFC works~\cite{flume2007,asbestos2005,histar2011}.
Hence, for intra-host communications, \tool uses eBPF programs to implicitly propagate labels by adding tags to the relevant BPF maps to satisfy the safe message rules. 
This allows the information to flow freely within a host according to the subject's choice.

For inter-host communications, \tool supports information declassification (or endorsement) controls by removing (or adding) tags in packets.
In existing OS-level DIFC systems, processes on hosts are responsible for declassifying or endorsing tags~\cite{flume2007}. However, this approach can potentially overwhelm the hosts,
especially for high-traffic networks, where the CPU can become a performance bottleneck when handling high-volume requests.
Therefore, \tool offloads the tag capabilities
to the high-speed programmable switches according to the defined \policy policies.
For a network flow that passes through the switch, the switch modifies the labeled packet header on the fly by removing secrecy tags (for declassification) or adding integrity tags (for endorsement).

The switch often needs to modify multiple tags at once.
For example, when a process exports data to the external world, the label for the network flow, potentially containing several tags, needs to be downgraded to an empty label $\{\}$.
To efficiently realize this, \tool uses \emph{bitmasks}. 
When an in-network policy is hit, the switch receives a bitmask {\cmtt mask} indicating which tags to modify (bit set to 1).
For declassification, the switch performs a bitwise AND operation between the DIFC packet header and $\sim${\cmtt mask} to clear secrecy tags.
For endorsement, the switch performs a bitwise OR operation with {\cmtt mask} to add integrity tags.

\begin{figure}[t]
\centering
\includegraphics[width=\linewidth]{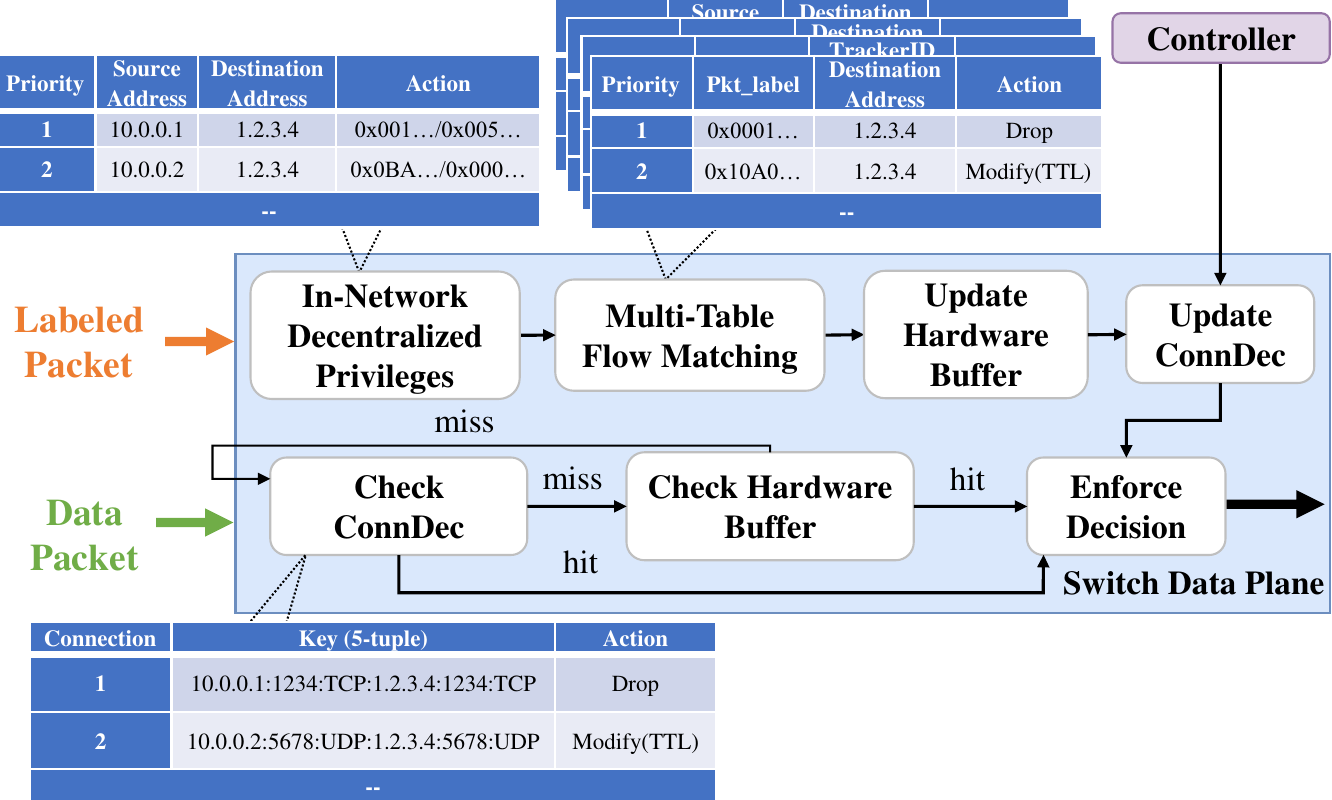}
\caption{In-network packet processing workflow}
\label{fig:p4arch}
\vspace{-1ex}
\end{figure}

\subsection{In-Network Policy Enforcement}
\label{subsec:policy_enforcement}

\policy policies defined by the network administrator are compiled into different \emph{in-network policies} to be executed in the switch.
For example, one type of in-network policy performs DIFC label pattern matching, which matches the network flow by examining the specific DIFC tags contained in the flow's DIFC packet header.
Other types of policies include matching by the flow's source and destination hosts and matching by the \texttt{TrackerID} tag.
To store these in-network policies in a switch, a naive way is to use a single large match/action table, similar to the traditional firewall structure. However, this design is highly inefficient as the policies that perform DIFC label pattern matching require ternary matching, which is expensive for DIFC packet headers of 32 bytes~\cite{ternarymatching} and must be placed in TCAM. Since TCAM has a much smaller capacity than SRAM, placing all policies in a single table in TCAM would quickly exhaust its capacity, resulting in only a few hundred policies that can be stored.

To store these policies efficiently within limited switch memory, \tool employs a \emph{multi-table flow matching} technique. 
Our idea is to place in-network policies in multiple match/action tables in different types of memories based on the type of matching. While policies that need DIFC label pattern matching are placed in TCAM, the other policy types that need exact matching are placed in the match/action tables in SRAM.
Our evaluation result in \cref{subsec:eval_overhead} shows that this approach increases the policy storage capacity by 12$\times$ compared to the single-table design.

\paragraph{Priority-based enforcement}
Inspired by the firewall policy design, the network administrator can define the priority of \policy policies, and \tool maintains the priority of each policy in its table entry. For example, in \cref{fig:motivatingexample_p4control}, the {\cmtt drop} policy (order~1) has a higher priority than the {\cmtt allow} policy (order~2). If a network flow matches multiple entries across different match/action tables, the switch will execute the one with the highest priority. If none of the policies match, the network flow will be dropped by default, resembling the default deny action of firewalls. 

\paragraph{In-network packet processing workflow}
As illustrated in \cref{fig:p4arch}, labeled packets (which carry a DIFC packet header) undergo initial matching against declassification/endorsement policies for tag modifications, followed by matching against in-network policies for a security decision. The decision is then stored in the hardware buffer and the {\cmtt ConnDec} table in the data plane. Subsequent packets in the network flow (which do not carry a DIFC packet header) are processed against the pre-determined decision stored in the data plane, ensuring quick handling of subsequent packets.

\subsection{Lightweight eBPF-Based Host Agent} \label{subsec:hostagent}

Our host agent persists DIFC context both within the host and from/to the network with minimal overhead. 
Enabled by eBPF (Extended Berkeley Packet Filter)~\cite{ebpf}, our host agent is lightweight and readily deployable without any kernel modifications. This differentiates our approach from previous DIFC works~\cite{flume2007,asbestos2005, histar2011}, which require extensive kernel modifications to track intra-host information flows.
eBPF is an emerging kernel technology that enables sandboxed programs to run in the kernel space \emph{without} modifying the kernel source code or loading additional modules. It enhances the performance, security, and flexibility of the kernel by allowing dynamic and event-driven programming.

\begin{figure}[t]
\centering
\includegraphics[width=\linewidth]{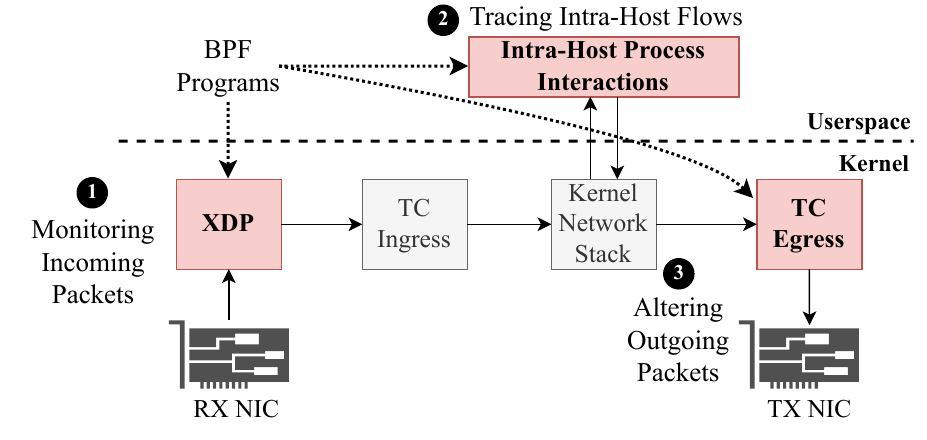}
\caption{eBPF hooks and intra-host label propagation path}
\label{fig:clientarch}
\vspace{-1ex}
\end{figure}

\paragraph{Lightweight DIFC label persistence}
Our host agent maintains lightweight label persistence by attaching carefully defined eBPF hooks in the kernel to capture the complete chain of intra-host events and accurately propagate DIFC labels.
This design makes our host agent compatible with a wide range of kernel versions, facilitating deployment in large, heterogeneous networks with various system configurations.
Fig.~\ref{fig:clientarch} illustrates these eBPF hook points. We create multiple BPF maps to share data between these eBPF programs.

\emph{\circled{1} Monitoring incoming packets:} 
The first hook monitors incoming labeled packets and extracts DIFC labels.
To achieve high-performance packet processing, we leverage the XDP (eXpress Data Path) \cite{xdp2018} technology to directly attach the eBPF program to the network device. When a new packet arrives, a callback invokes the eBPF program. If the ``evil bit'' of the packet is set, the eBPF program will extract the label information and store the destination port in a BPF map, {\cmtt inLabels (dport->[Label,TrackerID])}.
Then, the eBPF program removes the DIFC packet header and resets the ``evil bit'', restoring its original form for further kernel network stack processing. 
To identify which process receives the labeled packet, the eBPF program monitors processes that invoke a system call to receive a network connection. 
When a process accepts a connection, the eBPF program looks up the destination port in {\cmtt inLabels}.
If there is a match, the eBPF program will extract the label information and the process ID (PID) of the receiving process, and store the information in another BPF map, {\cmtt pidLabels} ({\cmtt PID->[Label,TrackerID]}).

\emph{\circled{2} Tracing intra-host flows:}
The second hook tracks the propagation of DIFC labels between processes and files during intra-host activities through a \emph{data provenance mechanism}.
This captures the chain of activities from the process that initially receives the labeled network flow to the process responsible for sending out network traffic.
The eBPF program monitors system calls related to process creation, and extracts the PID and the parent PID of the newly created process. 
Using the parent PID as a lookup key, the eBPF program retrieves the associated label from {\cmtt pidLabels} and propagates it to the child PID and updates {\cmtt pidLabels}.
If a process is terminated, the eBPF program will remove the entry from {\cmtt pidLabels}, ensuring that the same PID can be reused.
For file operations, the eBPF program maintains a BPF map, {\cmtt fileLabels} ({\cmtt Inode->[Label,TrackerID]}), that associates file inodes with their respective labels. When a new file is created, it is assigned the same label as the creating process, and {\cmtt fileLabels} is updated. This ensures that the file inherits the appropriate label and aligns with the security context of the creating process. During a file read, the eBPF program retrieves the file’s label from {\cmtt fileLabels} and uses it to update the label of the process (by updating {\cmtt pidLabels}) that performs the read operation. When a file is deleted, the eBPF program removes the entry from {\cmtt fileLabels} so that the inode can be reused. Tracking file activities helps identify attackers who may save stolen information in files and exfiltrate it later.

\emph{\circled{3} Altering outgoing packets:}
The third hook modifies the outgoing packets by incorporating the propagated DIFC labels. When a process invokes a system call to send a network message, the eBPF program is triggered to search for the PID in {\cmtt pidLabels}. Once the sending process is identified, the process's label and the source port are stored in a new BPF map, {\cmtt outLabels (sport->[Label,TrackerID])}. To match outgoing packets, we load the eBPF program into the TC (traffic control) Egress. When a packet exits, the eBPF program checks whether the packet's source port has been marked in {\cmtt outLabels}. If a match is found, the eBPF program will prepare the corresponding DIFC label in a DIFC packet header and insert the header into the outgoing packet.

\paragraph{BPF maps persistence}
As BPF maps reside in the kernel space, they are not persistent across eBPF programs reloading or system reboots. This can cause problems when the attacker performs file activities. Though the file data persists in the filesystem, the BPF maps can be lost, resulting in inaccurate label propagation. 

To keep BPF maps persistent across eBPF programs reloading, our host agent \emph{mounts} the eBPF virtual filesystem to the kernel memory. This allows the eBPF programs to pin their maps to the eBPF virtual filesystem by creating a file descriptor that points to these BPF maps. This file descriptor is linked to a specific pathname in the eBPF filesystem. As a result, the kernel will retain the BPF maps even if the referencing eBPF program is unloaded, as the corresponding file descriptor will keep pointing to the BPF maps. 

To further persist BPF maps across system reboots, our host agent \emph{migrates} the BPF maps to the permanent filesystem on the host machine.  When the host agent detects the {\cmtt kernel\_restart} or {\cmtt kernel\_power\_off} system events, it immediately migrates the BPF maps to a backup file in the permanent filesystem before the system reboots. After the system reboots, the host agent repopulates the BPF maps with the entries from the backup file. This repopulation occurs only once before the host agent resumes its functions upon reboot. 
Our host agent can also migrate BPF maps in case of a system crash, by detecting abnormal terminations of critical processes using the {\cmtt process\_exit} hooks.

\subsection{Distributed Multi-Switch Deployment}
\label{subsec:distributed}

\tool leverages the distributed nature of networks to optimize the deployment of in-network policies in the switches. 
A naive approach would install identical policies on every switch, which wastes space on switches that would never match those policies.
In contrast, \tool places policies only on switches that are likely to see the matching traffic.
This is achieved by placing each policy in the switch that is directly connected to the \emph{destination} address defined in the policy, similar to the setup of distributed firewalls. Once a flow is matched and is allowed to pass, the remaining switches only need to forward its packets, ensuring strict security and consistency of policy enforcement.
We choose the destination address instead of the source address because otherwise, an attacker could bypass the policies by using different hosts.
With this design, we can significantly reduce the storage overhead and minimize the unnecessary latency from re-matching the same flow.

\tool offers seamless integration into the existing network infrastructure that uses programmable switches, eliminating the need for installing additional middleboxes while offering minimal disruption to the network performance.
The central management of these switches by the control plane ensures up-to-date policy installation and simplifies maintenance.
Coordinating a distributed defense as a single entity in large infrastructures is complex.
Network segmentation~\cite{networksegmentation} addresses this challenge by dividing the network into distinct segments, each governed by its specific set of policies. \tool can be a pivotal facilitator in this design by configuring switches within each segment to manage their respective DIFC tags and policies. This configuration provides fine-grained control over individual segments, enhancing defense capabilities by accommodating a larger number of DIFC tags in the network.

\section{DIFC Policy Framework}
\label{sec:policy_language}

\tool provides an expressive DIFC policy language, named Network Control Language (\policy), for specifying diverse DIFC policies to counter different attack scenarios.
These policies are enforced in priority order and can be dynamically updated at runtime.
Despite multiple domain-specific languages proposed for network management~\cite{frenetic2011,Christopher2013} 
and network security~\cite{Vallentin2016,poseidon2020,qiao2020,psi2017}, none of them are designed for network-level DIFC policies.

\subsection{Expressive Policy Syntax}
\label{subsec:policysyntax}

\begin{figure}[t]
\centering
\includegraphics[width=.83\linewidth]{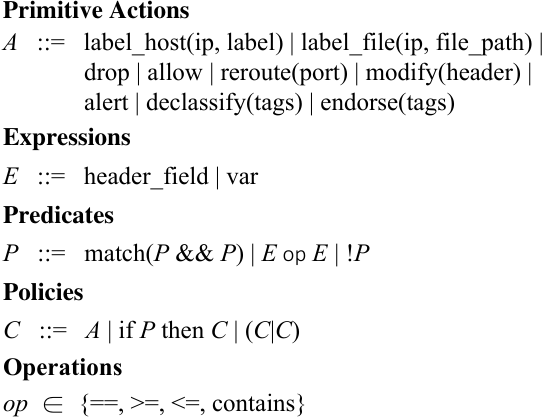}
\caption{Syntax of \policy}
\label{fig:policysyntax}
\vspace{-1ex}
\end{figure}

\cref{fig:policysyntax} illustrates the syntax of \policy, which is inspired by previous works \cite{poseidon2020, qiao2020} that adapt NetCore~\cite{Christopher2013} (an SDN programming language) for network defenses. 

\policy provides two labeling functions to initialize DIFC labels: 
{\cmtt label\_host(ip, label)} assigns a label to a specific host's IP address, initializing all existing processes and files with the host's label;
{\cmtt label\_file(host\_ip, file\_path)} assigns a unique \texttt{TrackerID} to a file on a host for fine-grained tracking of sensitive files. These functions can be used during initial deployment or subsequent stages when new hosts or files need labeling.

A \policy policy consists of a flow-matching predicate and an action. Various \emph{patterns} are provided to match a network flow based on the source and final destination hosts on a cross-host path, the DIFC tags, etc. 
Expressions {\cmtt (\emph{E})} can represent constants {\cmtt (var)} such as IP addresses and DIFC tags, as well as DIFC or IP packet header fields {\cmtt (header\_field)} such as {\cmtt dst\_ip} and {\cmtt pkt\_label}. 
\texttt{TrackerID} can be represented by the location of the tagged file (i.e., {\cmtt file\_path}@{\cmtt host\_ip)}.
Predicates {\cmtt (\emph{P})} are built over expressions with comparison operations {\cmtt (\emph{E op E})}, which are used to match network flows and trigger actions. The keyword {\cmtt contains} checks a subset of DIFC tags in the DIFC packet header of a network flow.

\policy provides multiple primitive \emph{security actions}. 
The {\cmtt drop} action discards a flow at the switch.
The {\cmtt allow} action forwards a flow based on the configured forwarding table.
The {\cmtt reroute(port)} action redirects suspicious traffic to a predefined destination, such as a logging server or a deep packet inspection (DPI) system, for further scrutiny or processing. 
The {\cmtt modify(header)} action uses the programmable parser of the switch to modify the packet header.
For instance, \tool can reset specific packet headers (e.g., IP options and TTL) which may be used as a covert channel for data exfiltration.
The {\cmtt alert} action serves as a detection mechanism, generating alerts to notify the network administrator when a suspicious flow is detected.

\policy also provides two \emph{privileged actions}. The {\cmtt declassify(tags)} action removes specified tags, allowing sensitive data declassification. The {\cmtt endorse(tags)} action adds designated tags, endorsing the flow's integrity.
Additionally, the endorsement action allows inserting tags for flows originating from external addresses, where no host agent is installed. 
This allows \tool to regulate information flows from external addresses within the network.

\subsection{Efficient Policy Compilation} \label{subsec:policy_compilation}

To enforce user-defined \policy policies in the data plane, \tool employs an efficient compiler to compile and execute \policy policies in the switch.
The label initialization statements are interpreted, and the switch uses its hardware packet generator to send a control packet containing the DIFC label to the respective host agent.

For \policy matching policies, it is essential to develop an efficient compilation strategy to counter the rapidly changing behaviors of attackers. When the attacker changes the strategies, the matching patterns must be updated accordingly, and the updated policy must be quickly recompiled and pushed to the switch.
A naive compilation strategy would compile a \policy policy into a P4 program to run in the switch. However, this approach would require reloading the P4 program every time a policy changes, which would interrupt the network traffic and cause significant disruption.

To improve defense agility, \tool employs an efficient compilation mechanism that supports \emph{dynamic update} of \policy policies without interrupting traffic.
Our idea is to compile \policy policies into the corresponding \emph{switch configurations}, which are a set of parameters that define a switch's operations, including match-action table entries for packet header matching, associated actions, and policy priority levels.
These switch configurations then insert in-network policies into their respective match/action tables within the switch.
Note that in this design, we still need to create a P4 program to specify all the logic to parse customized packet headers and define match/action tables.
However, this P4 program does \emph{not} contain specific match/action rules, and needs to be compiled by the P4 compiler (different from the \policy compiler) and loaded into the switch \emph{only once}. 
After the compilation, the switch configurations are passed to the switch daemon in the control plane.
Whenever the \policy policies change, the \policy compiler generates new switch configurations, and the switch daemon updates the match/action tables by adding or removing in-network policies accordingly.
This mechanism allows the control plane to seamlessly add or remove in-network policies.

\subsection{\policy Policy Examples}
\label{subsec:usecases}

We now present examples of \policy policies against various attack scenarios to show \policy's expressiveness.

\begin{figure}[t]
    \centering
    \includegraphics[width=.9\linewidth]{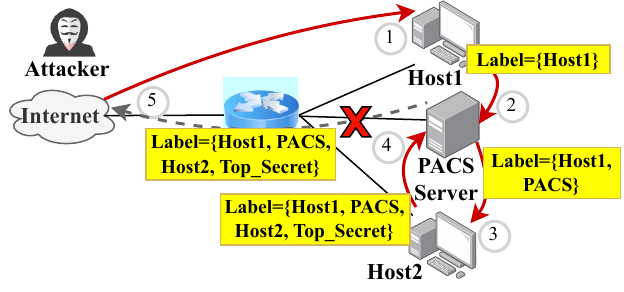}
    \caption{Preventing data exfiltration}
    \label{fig:usecase_DLP}
    \vspace{-1ex}
\end{figure}

\paragraph{Scenario 1: Preventing data exfiltration}
\cref{fig:usecase_DLP} illustrates a real-world data leakage incident against a hospital network~\cite{trapx}.
The attacker first compromises {\cmtt Host1}'s web browser to get into the internal network, aiming to exfiltrate sensitive data from {\cmtt Host2}.
The network has a picture archiving and communication system (PACS) server that is less secure and allows widespread data sharing. 
The firewall blocks direct connections from {\cmtt Host1} to {\cmtt Host2} and from {\cmtt Host2} to the external network. To bypass the firewall, the attacker uses the PACS server as a stepping stone to reach {\cmtt Host2} and then moves the data from {\cmtt Host2} to {\cmtt PACS} and ultimately to the external network.

\begin{lstlisting}[caption={Preventing exfiltration of top-secret data},label={lst:usecase_DLP}]
# Initialize labels
label_host(ip=Host1, label={Host1})
label_host(ip=Host2, label={Host2, Top_Secret})
label_host(ip=PACS, label={PACS})

# Drop network flows containing Top_Secret data
if match(pkt_label contains Top_Secret && dst_ip==external_network) then drop |\label{line:p1_drop}|

# Allow traffic between hosts and PACS server
if match(src_ip==Host1 && dst_ip==PACS) then allow
if match(src_ip==PACS && dst_ip==Host2) then allow
if match(src_ip==Host2 && dst_ip==PACS) then allow

... # Other policies that allow benign traffic
# DROP ALL (default deny)
\end{lstlisting}

\cref{lst:usecase_DLP} shows the \policy policies. We omit additional policies that allow benign traffic. 
Network flows that do not match any policies are dropped by default.
Using DIFC labels and a matching policy, we can protect {\cmtt Host2}'s data with the \texttt{Top\_Secret} tag from being leaked, regardless of intermediate hosts. For example, when the attacker attempts to export the secret data from {\cmtt PACS} to the external network, \tool detects the presence of the \texttt{Top\_Secret} tag in the network flow and blocks the flow
(\cref{line:p1_drop}).

\begin{figure}[t]
    \centering
    \includegraphics[width=\linewidth]{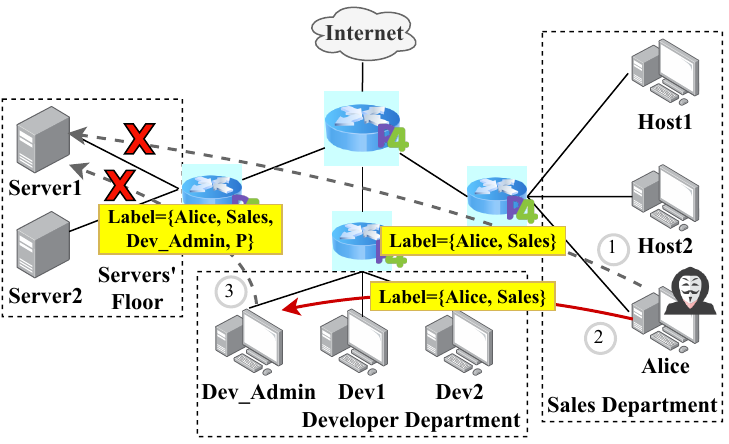}
    \caption{Preventing unauthorized access while endorsing high-integrity users to access}
    \label{fig:usecase_unauthorized}
    \vspace{-1ex}
\end{figure}

\paragraph{Scenario 2: Preventing unauthorized access}
Access control is essential in enterprise networks. However, existing solutions fall short of defending against insider threats across hosts.
This scenario showcases how we can leverage our in-network endorsement mechanism to prevent unauthorized access.
\cref{fig:usecase_unauthorized} illustrates an enterprise network where each department is protected by a firewall, only {\cmtt Alice} can access both the sales and developer resources, and only {\cmtt Dev\_Admin} has permission to access the Servers' Floor. Alice, as an insider attacker, can abuse her permissions and use a zero-day vulnerability to compromise {\cmtt Dev\_Admin} to gain further access to the Servers' Floor.

\begin{lstlisting}[caption={Endorsing users to access protected resources},label={lst:usecase_unauthorize}]
# Initialize labels
label_host(ip=Sales_Dept, label={Sales})
label_host(ip=Alice, label={Alice, Sales})
label_host(ip=Dev_Admin, label={Dev_Admin})

# Endorse network flows (add tag) from Dev_Admin
if match(src_ip==Dev_Admin && dst_ip==Servers_Floor) then endorse({P}) |\label{line:p2_endorce}|

# Only allow network flows with the integrity tag P
if match(src_ip==Sales_Dept && dst_ip==Servers_Floor) then drop |\label{line:p2_alice_block}|
if match(pkt_label contains P && dst_ip==Servers_Floor) then allow |\label{line:p2_check}|

... # Other policies that allow benign traffic
# DROP ALL (default deny)
\end{lstlisting}

\cref{lst:usecase_unauthorize} shows how we can prevent unauthorized insiders from accessing the Servers' Floor while only endorsing {\cmtt Dev\_Admin} for access. We only allow {\cmtt Dev\_Admin} to access the servers, by adding the tag \texttt{P} to flows originating from {\cmtt Dev\_Admin} (\cref{line:p2_endorce}) and checking the presence of the tag~\texttt{P} (\cref{line:p2_check}). 
Even if Alice exploits a vulnerability in {\cmtt Dev\_Admin} (\circledwhite{2}) to acquire the needed tag, the policy at \cref{line:p2_alice_block} will detect the \texttt{Sales} tag in the network flow and block Alice from connecting to the servers (\circledwhite{3}).

\begin{figure}[t]
    \centering
    \includegraphics[width=\linewidth]{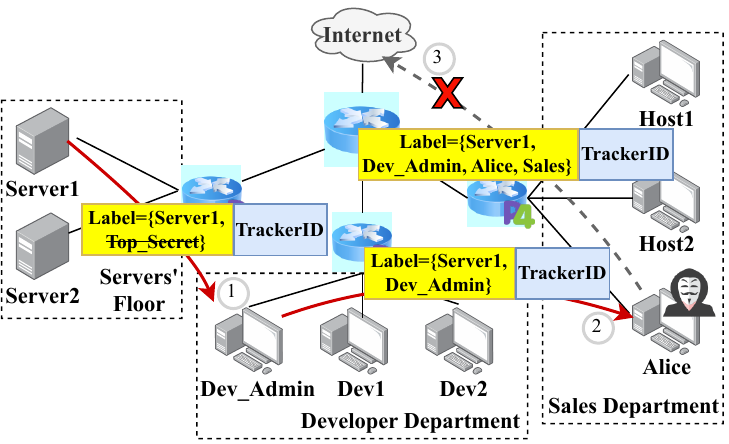}
    \caption{Fine-grained tracking of sensitive information}
    \label{fig:usecase_tracking}
    \vspace{-1ex}
\end{figure}

\paragraph{Scenario 3: Fine-grained tracking of sensitive information}
\cref{fig:usecase_tracking} illustrates the situation where fine-grained tracking is needed to further restrict the propagation of declassified files.
A protected file on {\cmtt Server1} is declassified to {\cmtt Dev\_Admin} for sharing within the company's internal network.
However, Alice may profit from gaining early access to confidential information and leaking the file to the external network without the permission of the company. 
\begin{lstlisting}[caption={Preventing exfiltration of declassified information},label={lst:usecase_tracking}]
#Initialize labels and TrackerID for the sensitive file
label_file(ip=Server1, file=/server1/sensitive_file) |\label{line:p3_label_file}|
label_host(ip=Server1, label={Server1, Top_Secret})
label_host(ip=Dev_Admin, label={Dev_Admin})
label_host(ip=Alice, label={Alice, Sales})

#Declassify Top_Secret (remove tag) data to Dev_Admin
if match(src_ip==Server1 && dst_ip==Dev_Admin) then declassify({Top_Secret}) |\label{line:p3_declassify}|

# Prevent the tainted file from leaving the network
if match(tracker_id==/server1/sensitive_file@Server1 && dst_ip==external_network) then drop |\label{line:p3_drop_taint}|

# Prevent Top_Secret data from leaving Server1
if match(pkt_label contains Top_Secret && dst_ip==any) then drop |\label{line:p3_drop_top_secret}|

... # Other policies that allow benign traffic
# DROP ALL (default deny)
\end{lstlisting}

\cref{lst:usecase_tracking} shows how we can track the propagation of sensitive information and prevent the information from being leaked to the external network. 
We assign a unique \texttt{TrackerID} tag to the sensitive file being tracked (\cref{line:p3_label_file}).
\texttt{Top\_Secret} files are protected and cannot leave {\cmtt Server1} without explicit declassification (\cref{line:p3_drop_top_secret}).
When \tool declassifies this file for {\cmtt Dev\_Admin}, it removes the \texttt{Top\_Secret} tag (\cref{line:p3_declassify}) but retains the \texttt{TrackerID} with the network flow (\circledwhite{1}). 
After Alice acquires a copy of the file (\circledwhite{2}) and attempts to leak it, the \texttt{TrackerID} persists with her outgoing network flow (\circledwhite{3}), enabling \tool to block the flow (\cref{line:p3_drop_taint}).

\begin{figure}[t]
     \centering

    \begin{subfigure}[b]{0.6\linewidth}
        \centering
        \includegraphics[width=\linewidth]{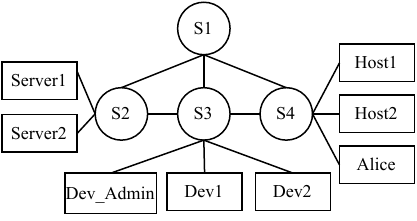}
        \caption{Our example enterprise network}
        \label{fig:enterprise}
    \end{subfigure}

    \vspace{4ex}
     
    \begin{subfigure}[b]{\linewidth}
        \centering
        \includegraphics[width=\linewidth]{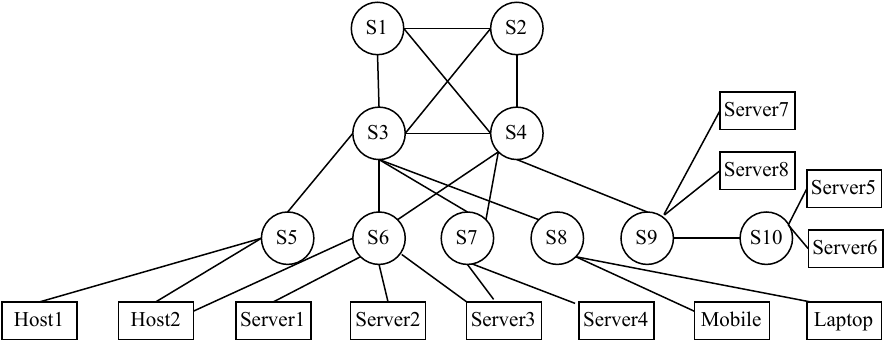}
        \caption{Cisco enterprise network}
        \label{fig:cisco}
    \end{subfigure}

    \vspace{4ex}

    \begin{subfigure}[b]{1\linewidth}
        \centering
        \includegraphics[width=\linewidth]{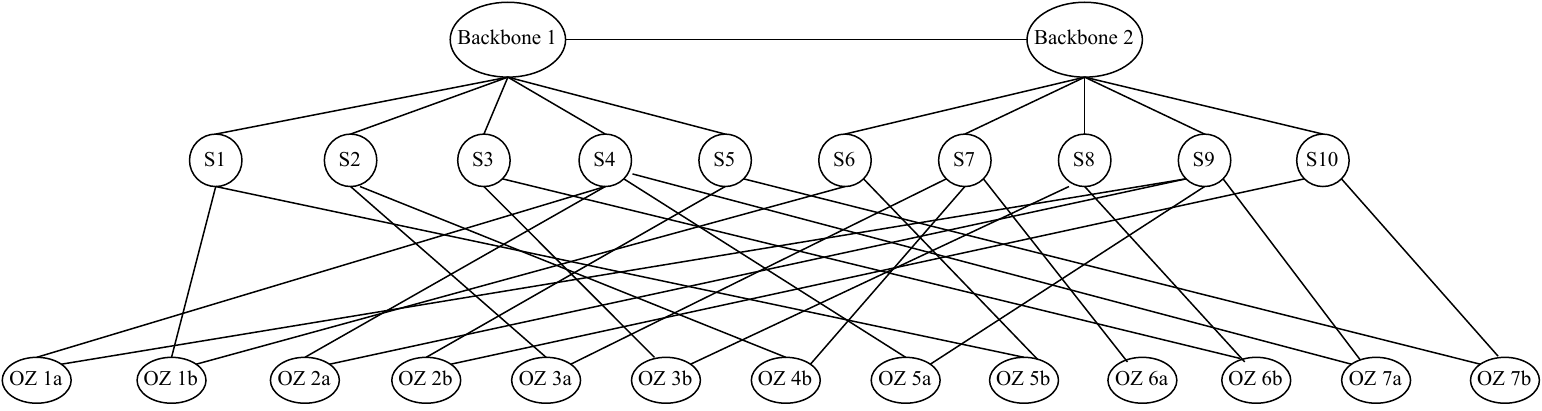}
        \caption{Stanford backbone network}
        \label{fig:stanford}
    \end{subfigure}
    \vspace{-1ex}
    \caption{Three network topologies (\emph{S} indicates switches)}
    \label{fig:topologies}
    \vspace{0ex}
\end{figure}

\section{Evaluation}
\label{sec:eval}

In this section, we aim to answer several key research questions on the defensive effectiveness and coverage, scalability, system capacity, and system overhead of \tool, through extensive evaluations.

\begin{enumerate}[label={\textbf{(RQ\arabic*)}}, leftmargin=*,itemsep=4pt, topsep=5pt]
  \item How effective is \tool in defending against various types of cross-host attacks?
  \item How well does \tool scale with real-world workloads and network topologies?
  \item What is the capacity of \tool and how does \tool impact the network performance and the host performance?

  \item How does \tool compare with existing SDN-based defenses?

  \item How does \tool's distributed deployment further optimize its efficiency and scalability?
\end{enumerate}

\paragraph{Implementation and deployment}
We implemented a prototype of \tool in $\sim$3,200~lines of P4, C, and Python code.
This includes the \policy compiler, eBPF programs, switch program, and switch control plane functions.
We deployed \tool on a testbed of a physical Wedge 100BF-32X Tofino P4 switch with 32×100~Gbps ports. The testbed setting is similar to that of existing P4 security works~\cite{Jiarong2020, jaqen2021}.
Our experiments run on three Dell R420 servers, each equipped with an Intel Xeon E5-2430 CPU running at 2.20~GHz, 64~GB RAM, and Ubuntu 20.04. 
We set up these three servers to act as the protected, intermediate, and attacker hosts, respectively.
Additionally, we deployed \tool in three representative enterprise topologies, including our example enterprise topology, Cisco enterprise network~\cite{psi2017}, and Standford backbone network~\cite{stanford_topology2012} (see \cref{tab:topologies} and \cref{fig:topologies}).
The topologies are constructed using a packet-level simulation in Mininet, which is integrated with a software P4 switch (i.e., bmv2~\cite{bmv2}) 
that emulates the behavior of physical switches.

\begin{table}[t]
\caption{Network topologies}
\centering
\resizebox{1\linewidth}{!}{
\begin{tabular}{@{}cccccccc@{}}
\toprule
\textbf{\Large Topology} & \textbf{\Large \# Hosts} & \textbf{\Large \# Switches} & \textbf{\Large Details} \\ 
\midrule
\textbf{\Large Example enterprise} & \Large 8 & \Large 4 & \Large Our example enterprise topology (\cref{fig:enterprise}) \\
\midrule 
\textbf{\Large Cisco} & \Large 14 & \Large 8 & \Large Cisco enterprise network (\cref{fig:cisco})\\ 
\midrule 
\textbf{\Large Stanford} & \Large 56 & \Large 25 & \Large Stanford backbone network (\cref{fig:stanford}) \\ 
\bottomrule
\end{tabular}}
\label{tab:topologies}
\vspace{-1ex}
\end{table}

\subsection{RQ1: Defense Effectiveness and Coverage} 
\label{subsec:eval_effectiveness}

We compare \tool with two real-world network defense solutions: a 
firewall (e.g., iptables~\cite{iptable}) and an NIDS (e.g., Snort~\cite{snort}). 
We configure these defenses to restrict the protected host from initiating or accepting connections from the external network (i.e., the attacker host), while the intermediate host is allowed to communicate with the other two hosts. 
This mirrors realistic enterprise setups where defenses shield vital resources from the attacker, with less restrictive policies on other devices for easier access.
To quantify the network performance, we measure the \emph{TCP congestion window} using iperf3,
which reflects the overall throughput.
We also measure \tool's impact on benign traffic by comparing the \emph{flow completion time (FCT)}  
under the typical forwarding switch (our baseline) and \tool defense. 
Background 
traffic is generated using the Distributed Internet Traffic Generator (D-ITG)~\cite{ditg}. 

We conduct nine attacks, categorized into two groups:
a stealthy insider attacker exfiltrating sensitive information (S1), and an external APT attacker accessing unauthorized resources (S2). 
In S1, we mimic an insider attacker residing in the protected host
and use netcat to transmit a stolen file to the intermediate host.
To exfiltrate the file to the attacker host, we use the Data Exfiltration Toolkit (DET)~\cite{det} with seven different communication channels, including different protocols and real-world applications (TCP, UDP, DNS, ICMP, HTTP, Twitter, and Gmail).
In S2, we use Metasploit from the attacker host to exploit a vulnerability~\cite{distcc} in the intermediate host, escalating privileges and establishing a connection to the protected host. We repeat the attack with the socat tool, which relays the attacker's traffic to the protected host from the intermediate host.

\begin{figure}
     \centering
    \begin{subfigure}[b]{0.49\linewidth}
    \centering
    \includegraphics[width=\linewidth]{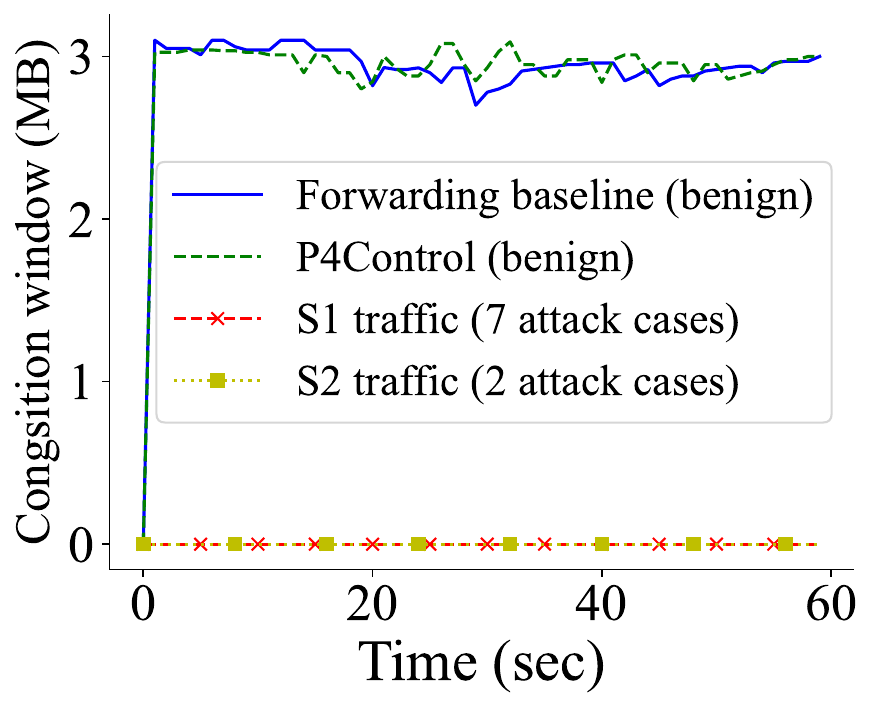}
    \caption{Congestion window}
    \label{fig:s1_s2_eval}
    \end{subfigure}
     \hfill
     \begin{subfigure}[b]{0.49\linewidth}
        \centering
        \includegraphics[width=\linewidth]{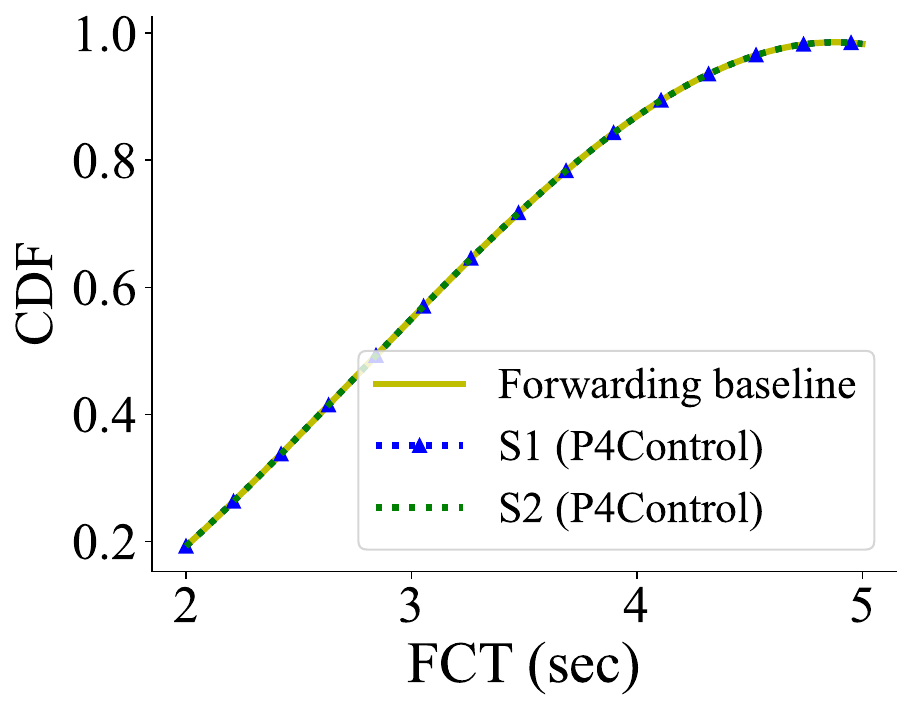}
    \caption{Benign traffic FCT}
    \label{fig:s1_s2_cdf}
     \end{subfigure}
        \caption{\tool blocks all stealthy cross-host attacks while imposing minimal overhead on benign traffic}
        \label{fig:datasets_cdf}
\end{figure}

Since the protected host and the attacker host do not communicate directly, both firewall and NIDS fail to block any attack.
With \tool, we label the protected host and the attacker host
and create a \policy policy to block communications between the two hosts.
As shown in \cref{fig:s1_s2_eval}, \tool blocks all nine attacks (i.e., congestion window at 0), ensuring security while maintaining a congestion window of benign traffic similar to the baseline.
Also, we observe no significant difference in the cumulative distribution function (CDF) of the FCT of the benign traffic between \tool and the baseline in \cref{fig:s1_s2_cdf}, showing that the defense has a negligible impact on benign traffic.

\subsection{RQ2: Scalability in Real-World Scenarios}
\label{subsec:eval_scalability}

\paragraph{Scalability with real-world enterprise workloads}
We evaluate the scalability of \tool using 
the Los Alamos National Laboratory (LANL) Unified Host and Network dataset~\cite{lanl} and the DARPA Operationally Transparent Cyber (OpTC) dataset~\cite{darpa}. 
The LANL dataset contains benign activities from 17,500 hosts.  
The DARPA OpTC dataset contains both benign and APT activities (initial compromise, lateral movements, privilege escalations, etc.) across 1,000 hosts.
Notably, in LANL, up to 80\% of daily network flows in enterprises use TCP protocol, emphasizing the need for efficient defenses that do not impact the sending rate.
Also, attackers typically pivot to infect more machines. 
In DARPA OpTC, an attacker pivoted across 14 hosts beyond its initial comprise, necessitating the need to limit the attacker's reachability.
Importantly, a very small portion of these enterprise events are associated with attackers' activities (e.g., only 0.0016\% in DARPA OpTC), 
requiring controls that do not interfere with benign traffic.

We use D-ITG to replay the two workloads to the physical switch and observe network performance during a ``no-defense'' baseline and under \tool. 
As LANL lacks malicious traces, we simulate real cross-host malicious flows (similar to S1 and S2 scenarios) for a comprehensive evaluation.
We label the initial victim hosts with a \texttt{V} tag and the final target hosts with a \texttt{protected} tag.
To block the multi-hop malicious access, we create the \policy policy: \incode{if match(pkt\_label contains V \&\& dst\_ip==target) then drop}.

By only appending DIFC labels to the SYN packet in TCP flows, our in-network per-flow decision technique effectively reduces the storage overhead of appending DIFC packet headers (6.4\% reduction for 500-byte data packets).
Additionally, \tool blocks all malicious network flows from reaching the target hosts regardless of 
the number of intermediate hosts compromised.
As shown in \cref{fig:datasets_cdf}, 
the ``no-defense’’ baseline and \tool have approximately the same average FCTs across all
flows in both datasets. 
This confirms that \tool scales well with real-world workloads, provides an effective defense against real-world cross-host attacks, and imposes minimal overhead on benign traffic due to its data plane enforcement.
Such observation is consistent with our testbed experiments in \cref{subsec:eval_effectiveness}.

\begin{figure}
     \centering
    \begin{subfigure}[b]{0.49\linewidth}
    \centering
    \includegraphics[width=\linewidth]{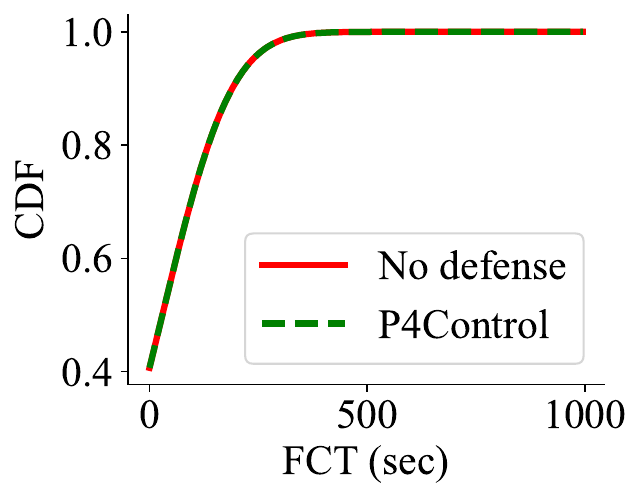}
    \caption{FCT in LANL}
    \label{fig:lanl_cdf}
    \end{subfigure}
     \hfill
     \begin{subfigure}[b]{0.49\linewidth}
        \centering
        \includegraphics[width=\linewidth]{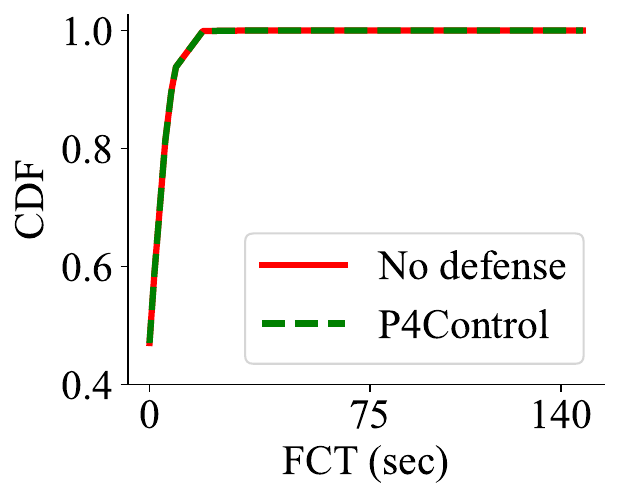}
        \caption{FCT in DARPA OpTC}
        \label{fig:darpa_cdf}
     \end{subfigure}
        \caption{\tool imposes minimal overhead under real-world enterprise workloads while blocking all attacks
        }
        \label{fig:datasets_cdf}
\end{figure}

\paragraph{Limiting attacker's reachability}
We analyze the reachability of an external APT attacker, who aims to 
access {\cmtt Server1} and {\cmtt Server2} in our enterprise topology in \cref{fig:enterprise}. 
Assuming that all hosts are vulnerable, we assess the attacker's reachability to the servers by exploiting each host and pivoting using varying \emph{step-counts}, which refers to the number of intermediate hosts used for pivoting.
In one setup, we configure the distributed firewalls as follows:
only hosts directly connected to switch {\cmtt S4} can access {\cmtt Server1}, only {\cmtt Dev\_Admin} can access {\cmtt Server2}, and hosts on {\cmtt S4} are blocked from those on {\cmtt S3}, except for Alice.
In another setup, we run \tool and assign a unique label to each host.

We run a script that simulates an external attacker performing APT steps (port scanning, exploiting hosts, escalating privileges, pivoting, etc.). 
The script aims to infiltrate hosts in the network and pivot towards the target servers, until either reaching them or exhausting the allowed step-counts.
\cref{tab:security_analysis} shows that the distributed firewall can easily detect attempts with step-count of 1, as they are direct accesses.
However, as the allowed step-count increases, the number of hosts that can be exploited and ultimately reach the servers increases. 
In contrast, \tool blocks all attempts 
regardless of the number of intermediate hosts.

\paragraph{Attack routes coverage}
To further evaluate \tool's attack routes coverage, we simulate an insider threat within the three enterprise networks in \cref{fig:topologies}.
We select a target machine and an insider attacker attempting to traverse the network to reach the target.
A firewall is configured to restrict the target's direct communications to a subset of allowed hosts inside the network.
We then run an attacker script that probes all possible routes to the selected target and records the number of successful accesses. 

\cref{tab:security_analysis2} shows (1) the number of potential attack routes under different network sizes and allowed step-counts and (2) attack routes coverage under different numbers of allowed hosts. 
Out of all potential attack routes to a target, the deployed firewall may only be able to block some of them. \emph{Attack routes coverage} refers to the proportion of attack routes that are blocked. The more hosts that are allowed direct access to the target, the lower the percentage of attack routes the firewall will be able to block.
This is common in enterprise networks, where a single user's access to multiple domains increases the potential attack routes. 
In contrast, \tool covers 100\% of attack routes and limits the target's access to the allowed hosts only. 
This holds no matter whether the attacker leverages the allowed host as a stepping stone to reach the target or not.
Notably, this protection is achieved using only a single \policy policy that blocks the attacker's label from reaching the target.

\begin{table}[t]
\caption{Number of hosts that can reach {\cmtt Server1} and {\cmtt Server2} with different step-counts in the example enterprise network in \cref{fig:enterprise}
}
\centering
\resizebox{0.75\linewidth}{!}{%
\begin{tabular}{@{}c|ccccc@{}}
\toprule
\textbf{Defense system} & \textbf{Step-count} & \textbf{Server1} & \textbf{Server2} \\ 
\midrule
\multirow{3}{*}{\textbf{Distributed firewall}} & 1 & 3 & 1 \\
\textbf{} & 2 & 7 & 3 \\
\textbf{} & 3 & 7 & 7 \\
\midrule 
\multirow{3}{*}{\textbf{\tool}} & 1 & 3 & 1 \\
\textbf{} & 2 & 3 & 1 \\
\textbf{} & 3 & 3 & 1 \\
\bottomrule
\end{tabular}%
}
\vspace{3ex}
\label{tab:security_analysis}
\end{table}

\begin{table}[t]
\caption{Number of attack routes in each network}
\centering
\resizebox{\linewidth}{!}{
\begin{tabular}{c|cc||ccc}
\toprule
\multirow{2}{*}{\textbf{\Large Topology}}  &
\multirow{2}{*}{\textbf{\Large Step-count}} &
\multirow{2}{*}{\textbf{\Large Total routes}} &
\multirow{2}{*}{\textbf{\Large Allowed hosts}} &
  \multicolumn{2}{c}{\textbf{\Large Attack routes coverage} \rule{0pt}{3ex}} \\
\cline{5-6}
\rule{0pt}{4.5ex} 
                &   &         &   & \textbf{\Large Distributed firewall} & \textbf{\Large \tool}  \\
\midrule
\multirow{5}{*}{\begin{tabular}[c]{@{}c@{}}\textbf{\Large Example} \\ \textbf{\Large enterprise}\end{tabular}} & \LARGE 6 & \LARGE 5,040    & \LARGE 1 & \LARGE 85\% & \multirow{5}{*}{\textbf{\LARGE 100\%}} \\
                                                    & \LARGE 5 & \LARGE 2,520    & \LARGE 2 & \LARGE 70\% &                      \\
                                                    & \LARGE 4 & \LARGE 840     & \LARGE 3 & \LARGE 57\% &                      \\
                                                    & \LARGE 3 & \LARGE 210     & \LARGE 4 & \LARGE 42\% &                      \\
                                                    & \LARGE 2 & \LARGE 42      & \LARGE 5 & \LARGE 28\% &                      \\
\midrule
\multirow{4}{*}{\textbf{\Large Cisco}} & \LARGE 5 & \LARGE 154,440  & \LARGE 2 & \LARGE 84\%       & \multirow{3}{*}{\textbf{\LARGE 100\%}} \\
                & \LARGE 4 & \LARGE 17,160   & \LARGE 4 & \LARGE 67\%       &                      \\
                & \LARGE 3 & \LARGE 1,716    & \LARGE 6 & \LARGE 53\%       &                      \\
                & \LARGE 2 & \LARGE 156     & \LARGE 8 & \LARGE 38\%       &                      \\
\midrule
\multirow{3}{*}{\textbf{\Large Stanford}}    & \LARGE 4 & \LARGE 8,185,320 & \LARGE 10 & \LARGE 81\%       & \multirow{3}{*}{\textbf{\LARGE 100\%}} \\
                & \LARGE 3 & \LARGE 157,410  & \LARGE 20 & \LARGE 63\%       &                      \\
                & \LARGE 2 & \LARGE 2,970    & \LARGE 30 & \LARGE 45\%       &                      \\
\bottomrule
\end{tabular}}
\label{tab:security_analysis2}
\end{table}

\paragraph{Maximum number of active connections}
We leverage the P4 compiler to assess the maximum number of active connections that \tool can handle, as it rejects a program if it consumes more memory than available resources.  
By progressively increasing the number of active connections, we find that \tool can support more than 220K concurrent active connections. 
This surpasses the number of active connections found in Facebook frontend clusters, which ranges from 10K to 100K~\cite{silkroad2017}.

\subsection{RQ3: System Capacity and Overhead}
\label{subsec:eval_overhead}

\begin{table}[t]
\caption{Number of supported in-network policies with different DIFC packet header sizes}
\centering
\begin{tabular}{@{}c|cc@{}}
\toprule
\textbf{DIFC packet header size} & 
\textbf{\# DIFC tags} & \textbf{\# In-network policies} \\
\midrule
\textbf{1-byte} & 8 & 140K \\
\textbf{2-byte} & 16 & 140K \\
\textbf{4-byte} & 32 & 72K \\
\textbf{8-byte} & 64 & 48K \\
\textbf{10-byte} & 80 & 36K \\
\textbf{16-byte} & 128 & 24K \\
\textbf{32-byte} & 256 & 12K \\
\bottomrule
\end{tabular}
\label{tab:system_policies}
\end{table}

\paragraph{In-network policies and DIFC tags supported}
We measure \tool's capacity in handling various DIFC packet header sizes and in-network policies within a single switch, by progressively increasing the header size and policy count until the P4 compiler rejects the program.
With a single flow matching table, \tool can maintain less than 1K policies due to the limited TCAM size.
In contrast, our multi-table flow matching technique enables \tool to maintain $\sim$12K unique policies, even with the largest 32-byte DIFC packet header that supports 256 tags. 
\cref{tab:system_policies} shows the number of in-network policies as the DIFC packet header size grows.
For comparison, the Stanford backbone network requires $\sim$1,500 access control list (ACL) policies~\cite{stanford_topology2012}. 
This indicates that a single switch deployment can accommodate many more policies than those typically used in large real-world networks.

As for tags, the single switch capacity for 256 tags largely exceeds the minimum access control requirement (16 sensitivity classifications and 64 categories) by the U.S. Department of Defense~\cite{tcsec}. 
This can be further increased through network segmentation as discussed in \cref{subsec:distributed}. 
Additionally, the latest switch models (e.g., Tofino~2/3~\cite{tofino2,tofino3}) have 3$\times$ more resources than our current model,
supporting a larger number of tags and in-network policies.

\begin{figure}[t]
\centering
    \includegraphics[width=\linewidth]{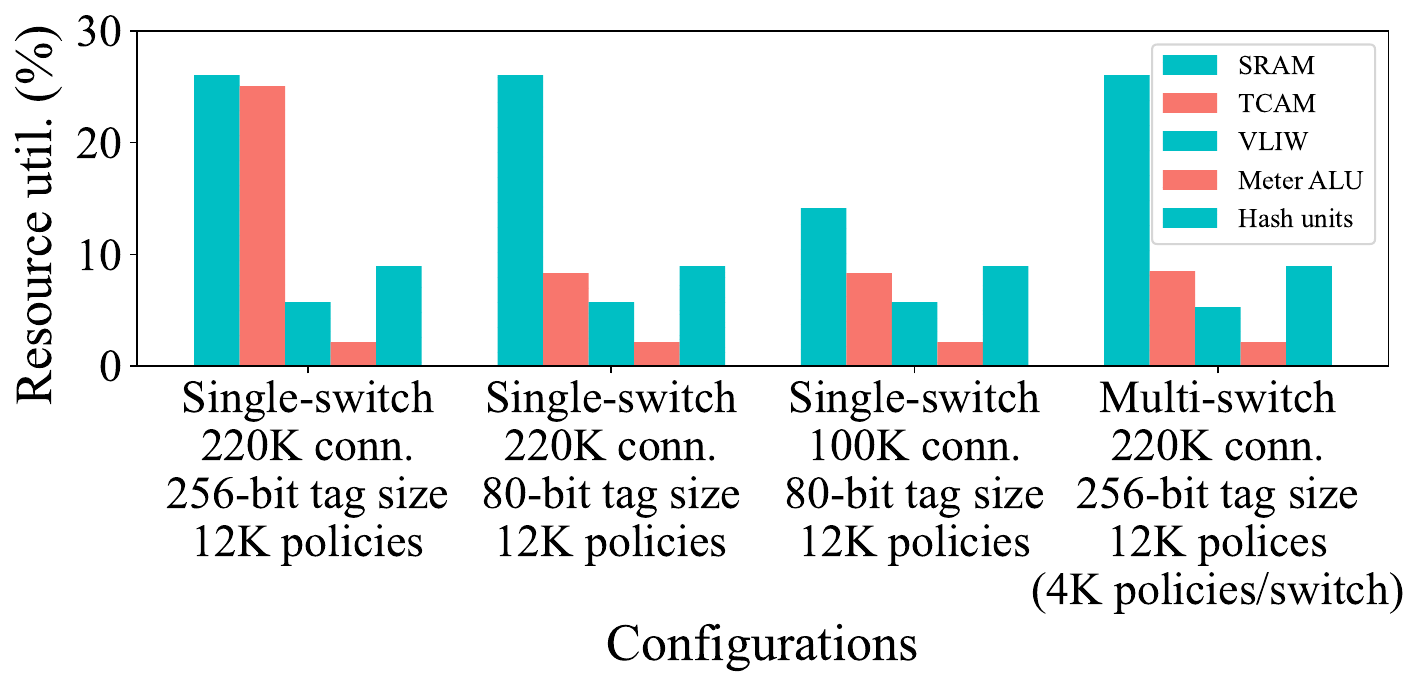}
    \caption{Switch resource utilization in different settings}
    \label{fig:resources}
\vspace{-1ex}
\end{figure}

\paragraph{Switch resource utilization}
\cref{fig:resources} shows how the switch resource utilization varies with the number of active connections, the tag size, and the number of policies.
In the single-switch deployment, with 220K active connections, 256-bit tag size, and 12K policies, the resource utilization is 26\% SRAM, 25\% TCAM, 5.7\% VLIW, 2.1\% meter ALU, and 8.9\% hash units. When we reduce the tag size to 80 bits while keeping the same number of active connections and policies, the TCAM utilization decreases to 8.3\%, while other resources remain roughly unchanged. This indicates that the tag size has a significant impact on the TCAM utilization, as matching larger DIFC packet headers requires more TCAM. We repeat the experiment but with 100K active connections, and the SRAM utilization decreases to 14.1\% while other resources remain unchanged. This indicates that the more active connections the switch handles, the more SRAM it requires for maintaining decisions of established connections.

We further compare the resource utilization of single-switch deployment and multi-switch deployment. We 
use 220K active connections, 256-bit tag size, and 12K policies, and deploy \tool on three switches to distribute the policies. As shown in \cref{fig:resources}, the TCAM utilization noticeably drops from 25\% to 8.5\% on each switch, as only 4K policies are maintained within each switch.

\begin{figure}[t]
 \centering
        \includegraphics[width=0.9\linewidth]{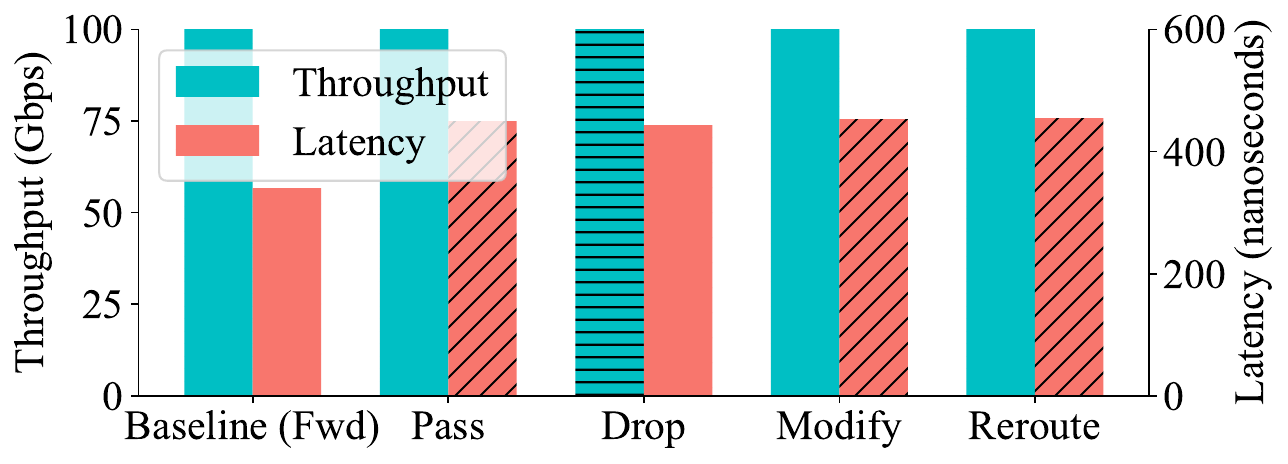}
        \caption{Throughput and latency under \tool compared to the forwarding baseline}
        \label{fig:system_overhead}
\vspace{0ex}
\end{figure}

\paragraph{Impact on network throughput and latency}
We evaluate the impact of \tool on the network throughput and latency. 
We compare different \tool actions with a forwarding baseline program (Fwd).
We use the on-switch hardware packet generator, which can generate 100~Gbps (per-port) traffic for stress testing.
As shown in \cref{fig:system_overhead}, \tool achieves a throughput of 99.9~Gbps, maintaining the highest performance of our switch. Also, it introduces a latency overhead of 100-110~ns when compared with the 
baseline, which is negligible since the RTT in typical enterprise networks is in the order of milliseconds~\cite{rtt2009}.

\paragraph{Host agent overhead}
We evaluate the host agent's overhead on a Linux server (kernel version 5.15.0) by measuring the average \emph{additional latency} of the eBPF programs over 10K runs.
For network ingress and egress eBPF programs, we generate 10K labeled network flows with 10 packets each, and measure the runtime of these two programs.
\cref{tab:overhead} shows that the host agent adds an overhead of 1-7 ms per system call, which is negligible compared to the total runtime of most system calls.  
Notably, for read/write operations, the eBPF programs take a similar time as the actual system call to record the label in the corresponding BPF map. 
However, they execute after the system call returns, thereby not blocking the completion of system call operations.

We further measure the storage overhead of maintaining DIFC labels using BPF maps. For {\cmtt pidLabels}, as there are only $2^{15}$ process IDs typically available in a Linux system, our table consumes a size of 1.3~MB, which is negligible considering the current abundant storage in systems. As for {\cmtt fileLabels}, the default setting in Linux allocates one inode for every 16~KB of space. Therefore, in a 1~GB filesystem, {\cmtt fileLabels} consumes 2.6~MB to maintain the labels for the inodes that can be assigned, which only takes 0.25\% of the whole storage in the filesystem.

\begin{table}[t]
\caption{Host agent overhead (in ms)}
\centering
\resizebox{\linewidth}{!}{%
\begin{tabular}{@{}ccccccc@{}}
\toprule
\textbf{} & \textbf{\Large execve} & \textbf{\Large clone}  & \textbf{\Large TC egress} & \textbf{\Large XDP ingress} & \textbf{\Large read} & \textbf{\Large write} \\ 
\midrule
\textbf{\Large System call time} & \Large 466 & \Large 266 & \Large 59 & \Large 36 & \Large 8 & \Large 9\\ 
\midrule
\textbf{\Large \tool overhead} & \Large +6 & \Large +5 & \Large +0.7 & \Large +0.3 & \Large +5 & \Large +6 \\
\bottomrule
\end{tabular}%
}
\label{tab:overhead}
\vspace{-1ex}
\end{table}

\subsection{RQ4: Comparison with SDN-Based Solution}
\label{subsec:eval_sdn}

We compare \tool with PivotWall~\cite{oconnor2018}, an OpenFlow SDN-based IFC approach. PivotWall propagates taint tags between system entities within hosts. When two hosts communicate, the sender host adds the taint tag to outgoing packets and sends ``control messages'' 
with the taint information to the SDN controller. The controller maintains a local graph of the taint propagation path.
Upon receiving the ``control messages'', the controller matches the incoming tainted network flow and the local graph with the policies. Then, the controller installs the corresponding decision in the switch. As PivotWall is not open-sourced, we implement it as an SDN application. In our testbed, we set up a Floodlight SDN controller on one server and configure another one with OpenvSwitch for OpenFlow communication. We use a third server to generate traffic for evaluation.

\paragraph{Precise information confinement}
PivotWall relies solely on the taint information of system entities and lacks precise information confinement. It also lacks safe controls to declassify or endorse data.
Furthermore, it uses a modified Linux kernel to track information within hosts, which requires intensive kernel changes.
In contrast, \tool supports a full DIFC model for precise information confinement via DIFC labels, and has mechanisms to safely move data between different compartments. This level of fine-grained control cannot be achieved in PivotWall with coarse-grained tracking. In addition, \tool can be seamlessly integrated with hosts using eBPF without kernel modifications, largely simplifying the agent deployment.

\paragraph{Defense responsiveness}
We compare the time taken by each defense to install the decision after receiving a tainted packet. 
With PivotWall, the ``control messages'' must be routed to the control plane for matching, incurring a \emph{round-trip delay} before a decision is pushed to the switch. 
Our measurement shows that this process takes between 8~ms to 3~seconds until a decision is pushed, depending on network traffic and graph lookup time. 
In comparison, \tool performs the flow matching in the \emph{data plane}, achieving a much smaller delay of less than 500~ns.

\begin{figure}[t]
\centering 
    \includegraphics[width=0.75\linewidth]{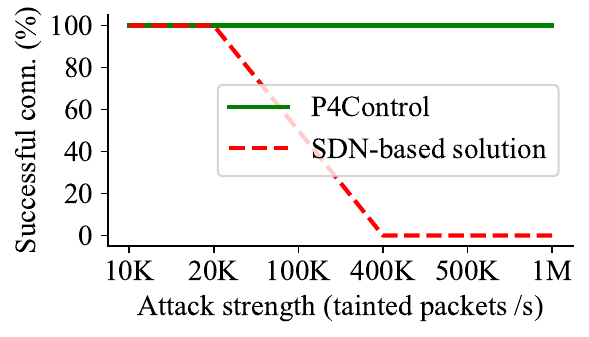}
    \caption{Control plane saturation attack against an SDN-based solution (PivotWall~\cite{oconnor2018}) and \tool}
    \label{fig:sdn}
\vspace{-1ex}
\end{figure}

\paragraph{Control plane saturation attacks}
Centralized SDN-based solutions introduce a single point of failure, potentially causing performance bottlenecks and security vulnerabilities.
We simulate an attacker that overwhelms the link between the switch and the controller by initiating a large number of connections.
\cref{fig:sdn} shows how PivotWall's central controller struggles to process new connections when the attack strength exceeds 100K tainted packets/s. At an attack strength of 1M packets/s, \emph{99\% of the legitimate connections are dropped}, making PivotWall ineffective. Since \tool examines packets entirely in the data plane, it maintains a stable performance and installs 100\% of the connections during the attack.


\subsection{RQ5: Distributed Storage Optimization}
\label{subsec:distributed_eval}

Our distributed 
multi-switch deployment significantly reduces the storage overhead of in-network policies.
Applied to our enterprise network in \cref{fig:enterprise}, with 100 in-network policies per host (amounting to 800 unique policies), 
\tool distributes the policies across the three switches that are directly connected to the hosts. 
Such distribution reduces each switch's storage overhead by an average of $\sim$66\%.
This allows for efficient utilization of the switch resources and ensures consistent policy enforcement.
Also, by performing the policy matching on a single switch, \tool only incurs an additional latency of $\sim$110~ns, which is negligible, regardless of the number of switches that a network flow traverses on its path to its destination.

\section{Discussion}
\label{sec:discussion}

\paragraph{Limitations of host agent} 
While eBPF provides a secure sandbox environment for running user-defined programs in the kernel, there are still some potential vulnerabilities with the current technology.
By exploiting existing vulnerabilities within eBPF through malicious code~\cite{ebpf-vul1},
an attacker could execute arbitrary memory reads and writes, compromising the integrity of our BPF maps.
Fortunately, there have been efforts to harden eBPF through improved safety verification of eBPF programs~\cite{unleashing2023} and fine-grained BPF privileges~\cite{bpfprivilages}, which can enhance the security 
of our host agent.

\paragraph{Robustness against integrity poisoning}
\policy policies can prevent integrity poisoning attacks that target DIFC systems.
In such attacks, malicious hosts with low integrity levels may attempt to connect to benign hosts with high integrity levels, lowering their integrity and restricting their access to high-integrity resources. 
Through \policy policies, the network administrator can assign appropriate integrity levels to hosts or domains, block low-integrity network flows from communicating with high-integrity hosts, and selectively endorse valid flows only from hosts with access permissions. 
With flexibility and expressiveness of \policy, the network administrator can safeguard communications across different integrity levels and block poisoning attempts.

\paragraph{Policy deployment in dynamic scenarios}
\tool reduces the switch resource utilization by distributing policies to multiple switches. However, this static deployment can face challenges in dynamic network environments, such as network topology changes or switch failures, which require policy reallocation. 
Moreover, the switch has limited resources that must be shared with other data plane applications.
The available resources may vary dynamically due to policy updates and the loading/unloading of other applications, which further complicates the policy deployment.
\tool can benefit from an online policy deployment strategy that dynamically reallocates policies while ensuring enforcement consistency and balancing resource usage.

\paragraph{Zero trust architecture}
Zero trust (ZT) is an evolving set of security paradigms that assume no implicit trust of any user account or asset based on their physical or network location or ownership. Instead, ZT requires persistent verification of every interaction with the least privileges granted~\cite{zerotrust}.
It is a radical shift from the traditional ``castle-and-moat'' network security model that relies on perimeter defenses and implicit trust inside the network. Motivated by the U.S. White House issued Executive Order EO-14028~\cite{order-14028} and Memo M-22-09~\cite{memo-m2209}, ZT has recently gained wide attention.
\tool's ability for fine-grained least-privilege network access control via in-network DIFC aligns with ZT principles.

\tool can be further extended to realize ZT goals in enterprise networks.
More types of complex security and integrity policies that incorporate behavioral host attributes can be designed. These attributes can be collected by our eBPF-based host agent and analyzed by an intelligent data plane that runs a machine learning model (e.g., decision tree~\cite{zhou2023}). 
This behavioral analysis enables continuous assessment of user and device profiles for adaptive access control. 
Being integrated into the existing network infrastructure with minimal modifications and overhead, \tool transforms the network into a defense backbone, serving as a valuable component for implementing a ZT architecture in enterprise networks.

\section{Related Work}
\label{sec:related_work}

\paragraph{Programmable switches} 
Recent works proposed to offload networking tasks to the data plane \cite{belma_fast2018, flowradar2016}. In addition, many works leverage data plane programmability to develop security primitives that run at line rate \cite{poseidon2020,jaqen2021,qiao2020,Jiarong2020,Jiarong2021,mew2023}.
Unlike \tool, none of these works focus on real-time prevention of sophisticated cross-host attacks.

\paragraph{DIFC}
In \cref{sec:motivation_background}, we reviewed existing DIFC works in detail, discussed their limitations, and explained why they are unsuitable for our goal. 
\tool is the first work that realizes DIFC at the network level at line rate.

\paragraph{System auditing}
Prior works proposed to collect system audit logs of system calls and construct system provenance graphs to aid attack investigation. These works, such as \cite{protracer2016, samuel_backtracking2003,nahid2017,fang2022back}, 
proposed different techniques for comprehensive system provenance analysis. Other works discussed cross-host attacks by associating host-level provenance~\cite{yangji2017, yangji2018, spade2012, nahid2020}, which primarily target post-attack forensic investigation instead of real-time attack prevention.
\tool differs from all these system-level defenses in proposing a new paradigm of network-level APT defenses using programmable data planes with line-rate defense performance.

Recent works also proposed domain-specific languages to query attack behaviors from system audit logs~\cite{gao2018aiql,gao2018saql,gao2021enabling,pasquier2018runtime}. However, these languages are not designed for network-level DIFC policies and are unable to express complex secrecy and integrity policies either.

\paragraph{SDN} Recent works proposed SDN-based solutions to extend packets with taint tags derived from host-level information \cite{oconnor2018}. 
However, their centralized design incurs high network latency and exposes additional attack vectors to the control plane. 
\tool leverages data plane programmability to address these issues, augmenting the defense with line-rate performance and minimal overhead.

\section{Conclusion and Future Work}
\label{sec:conclusion}

We proposed \tool, a network defense system for preventing cross-host attacks in real time.
\tool employs a novel in-network DIFC mechaism based on programmable switches and eBPF, and offers an expressive policy framework for specifying DIFC policies.
\tool is effective against various cross-host attacks while maintaining line-rate performance with minimal overhead.

There are a few future directions that are worth exploring.
First, \tool's DIFC enforcement scope can be extended to include the confinement of information within hosts. This can be achieved by extending the functionalities of our host agent, similar to previous OS-level DIFC systems, but with minimal kernel modifications and host overhead offered by eBPF.
Second, we can design a dynamic multi-switch deployment strategy using online optimizations, which can optimize the policy deployment based on available switch resources and adapt to dynamic network changes.
Third, we can extend \tool to implement a zero trust architecture, with more complex secrecy and integrity policies that incorporate behavioral host attributes.

\section*{Acknowledgement}
We would like to thank the anonymous reviewers and our shepherd for their constructive comments and suggestions.
This work is supported in part by the Commonwealth Cyber Initiative (CCI). Any opinions, findings, and conclusions made in this paper are those of the authors and do not necessarily reflect the views of the funding agencies.

\bibliographystyle{IEEEtran}
\bibliography{ref}


\begin{appendices}

\newpage 


\section{Meta-Review}

The following meta-review was prepared by the program committee for the 2024
IEEE Symposium on Security and Privacy (S\&P) as part of the review process as
detailed in the call for papers.

\subsection{Summary}
This paper proposes \tool, a network defense system capable of controlling flows and preventing cross-host attacks in real time by leveraging programmable switches and eBPF. \tool creates and propagates DIFC labels of network flows and acts on labeled flows in the data plane according to DIFC rules specified by the network administrator. The authors demonstrate that \tool is feasible for switch hardware, lightweight on host agents, and effective for various cross-host attacks.

\subsection{Scientific Contributions}
\begin{itemize}
\item Creates a New Tool to Enable Future Science
\item Addresses a Long-Known Issue
\item Provides a Valuable Step Forward in an Established Field
\item Establishes a New Research Direction
\end{itemize}

\subsection{Reasons for Acceptance}
\begin{enumerate}
\item This paper creates a new tool, \tool, which enforces DIFC from the network at line rate using programmable switches and eBPF. The approach further enables practical DIFC enforcement by using new networking architectures to implement it.

\item This paper presents a system that solves a well-motivated problem. Technical details on enforcing DIFC in the data plane are interesting and well written.

\item The proposed approach also establishes a new research direction in using programmable data planes to enforce complex integrity and security policies at line rate.

\end{enumerate}

\end{appendices}

\end{document}